\documentclass[12pt]{article}
\usepackage{amssymb}
\usepackage{amsmath}
\usepackage{amsthm}
\usepackage{cancel}
\usepackage{slashed}
\usepackage{fullpage}
\usepackage{color}
\usepackage{braket}
\usepackage{graphicx}
\usepackage[small]{subfigure}
\usepackage{multirow}
\usepackage{verbatim}
\usepackage{cite}
\usepackage{bigints}
\usepackage{simplewick}
\usepackage{authblk}
\usepackage{hyperref}
\hypersetup{
    linktocpage,
     colorlinks,
     citecolor=darkgreen,
     linkcolor= darkgreen,
     urlcolor=darkgreen
}

\def\cC{\mathcal{C}}

\def\cI{\mathcal{I}}

\def\cM{\mathcal{M}}

\def\cO{\mathcal{O}}

\def\cS{\mathcal{S}}

\def\imply{\quad\Longrightarrow\quad}

\def\dg{\dagger}

\newcommand{\Sl}[1]{\slashed{#1}}

\def\e{\varepsilon}
\def\be{\begin{equation}}
\def\ee{\end{equation}}

\def\conv{\otimes}

\newcommand{\Eq}[1]{Eq.~\eqref{#1}}

\newcommand{\SoftSub}[1]{{\color{blue} \overset{\substack{ #1 \\ \text{soft} \\ \text{sub} }}{} }}
\newcommand{\SoftSubNO}[1]{{\color{blue} \substack{ #1 \\ \text{soft} \\ \text{sub} } }}

\def\LPeq{\cong}
\def\LPFeq{\cong_{\text{\tiny{IR}}}}

\definecolor{darkred}{rgb}{0.5,0.0,0.0}
\definecolor{darkblue}{rgb}{0.0,0.0,0.9}
\definecolor{darkerblue}{rgb}{0.0,0.0,0.5}
\definecolor{darkgreen}{rgb}{0.0,0.5,0.0}
\definecolor{black}{rgb}{0.0,0.0,0.0}
\definecolor{brown}{rgb}{0.6,0.4,0.2}
\newcommand{\red}{\color{darkred}}

\newcommand{\ccol}{\color{darkblue}}

\newcommand{\softcol}{\color{darkred}}

\newcommand{\hch}{ {\text{h}} }
\newcommand{\scs}{ {{\softcol s}} }

\newcommand{\scL}{ {{\softcol \Lambda}} }
\newcommand{\scR}{ {{\softcol R }} }
\newcommand{\scRb}{ {{\softcol \overline R }} }
\newcommand{\scinf}{ {{\softcol \infty }} }

\newcommand{\ccO}{ {{\ccol 1}} }
\newcommand{\ccT}{ {{\ccol 2}} }
\newcommand{\cci}{ {{\ccol i}} }
\newcommand{\ccj}{ {{\ccol j}} }

\newcommand{\rN}{ {{\ccol N}} }

\newcommand{\ccN}{ {{\ccol N}} }
\newcommand{\ccc}{ {{\ccol c}} }
\newcommand{\ccR}{ {{\ccol R}} }

\newcommand{\eir}{ \varepsilon_{\text{IR}} }
\newcommand{\euv}{ \varepsilon_{\text{UV}} }

\newcommand{\SL}{ {\red {\cS}}}

\makeatletter
\newcommand*{\hyperlinkcite}[1]{\hyper@link{cite}{cite.#1\@extra@b@citeb}}
\makeatother
\newcommand{\Tree}{[\hyperlinkcite{Feige:2013zla}{FS1}]}
\newcommand{\Loop}{[\hyperlinkcite{Feige:2014wja}{FS2}]}

\newcommand{\Sij}{ S_{\cci \ccj}}

\newcommand{\IRZ}{\widehat{Z} }

\newcommand{\eik}{\text{eik}}

\def\rd{\textrm{d}} 
\newcommand{\wtd}{\widetilde}
\title{Avoiding double-counting in factorized cross sections}
\title{Removing phase-space restrictions in factorized cross sections}
\author{Ilya Feige\thanks{feige@physics.harvard.edu}}
\author{Matthew D. Schwartz\thanks{schwartz@physics.harvard.edu}}
\author{Kai Yan\thanks{kyan@physics.harvard.edu}}

\affil{\emph{Center for the Fundamental Laws of Nature,
Harvard University, Cambridge, MA 02138, USA}}

\begin{document} 
\maketitle

\begin{abstract}
Factorization in gauge theories holds at the amplitude or amplitude-squared level for states of given soft or collinear momenta. When
performing phase-space integrals over such states, one would generally like to avoid putting in explicit cuts to separate soft from collinear momenta.
Removing these cuts induces an overcounting of the soft-collinear region and adds new infrared-ultraviolet divergences in the collinear region. 
In this paper, we first present a regulator-independent subtraction algorithm for removing soft-collinear overlap at the amplitude level which may be
useful in pertubative QCD.
We then discuss how both the soft-collinear and infrared-ultraviolet overlap can be undone for certain observables in a way which respects factorization. 
Our discussion
 clarifies some of the subtleties in phase-space subtractions and 
includes a proof of the infrared finiteness of a suitably subtracted jet function.
 These results complete the connection between factorized QCD and Soft-Collinear Effective Theory.
\end{abstract}

\newpage

\section{Introduction}

Factorization is at the heart of our ability to use perturbative quantum chromodynamics (QCD) to make theoretical predictions for scattering processes at high-energy particle colliders. It is extremely fortuitous that accurate particle distributions can be computed by convolving universal parton distribution and hadronization models with perturbative calculations of jet formation. While factorization at the non-perturbative level is hard to establish, factorization relevant to the structure and substructure of jets can be understood within perturbation theory. In particular, the radiation patterns in perturbative QCD factorize into hard, collinear and soft contributions.
Moreover, subtleties in perturbative factorization (for example, related to non-global logarithms~\cite{Dasgupta:2001sh,Banfi:2002hw,  Kelley:2011tj,Schwartz:2014wha,Larkoski:2015zka,Caron-Huot:2015bja,Khelifa-Kerfa:2015mma}) are a limiting factor in many ultra-precise jet-substructure calculations. Thus, there has recently been renewed interest in studying factorization, particularly in the context of Soft-Collinear Effective Theory (SCET).

A concise formulation of factorization in QCD was proposed and proven in \cite{Feige:2013zla} and \cite{Feige:2014wja}, hereafter referred to as \Tree{} and \Loop{} respectively. These papers build upon decades of insight~\cite{Bloch:1937pw,Weinberg:1964ew,Coleman:1965xm,Sterman:1978bi,Collins:1981uk,Bauer:2001yt,Freedman:2011kj}. Up to color factors, the formula from~\Loop{} reads:
\be
\bra{X_\ccO\cdots X_\ccN; X_\scs}\cO\ket{0}
 \;\LPeq\; 
\cC_\cO(\Sij) \, \frac{\bra{X_\ccO} \bar\psi W_\ccO \ket{0}}{\bra{0} Y_\ccO^\dg W_\ccO \ket{0}} \,\cdots\, \frac{\bra{X_\rN} W_\rN^\dg\psi \ket{0}}{\bra{0} W_\rN^\dg Y_\rN \ket{0}}
\;\bra{X_\scs} Y_\ccO^\dg \cdots Y_\rN \ket{0}
\label{QEDmain}
\ee
In this expression, the state $\bra{X_\ccO\cdots X_\ccN;  X_\scs}$ has soft particles, in $\bra{X_\scs}$, and particles collinear to various specified directions, in $\bra{X_\cci}$. The left-hand side is a matrix element in QCD of an operator like $\cO = \bar{\psi}\cdots \psi$ in this state. The right hand side is a factorized product of matrix elements, each of which involves only one collinear sector or the soft sector. The symbol $\LPeq$ indicates that the two sides are identical at leading power. More precisely, if one were to compute some infrared-safe observable dominated by soft or collinear radiation, such as the sum of the jet masses $\tau=\frac{1}{Q^2}\sum{m_i^2}$, all of the terms in $\frac{d\sigma}{d\tau}$ that are dominant as $\tau \to 0$ will be
identical on both sides. 
More details can be found in Section~\ref{sec:AmpLevel} below and in \Tree{} and \Loop{}.

The formula in Eq.~\eqref{QEDmain} presupposes that the external momenta are designated as soft or collinear. If a particular momentum can be classified as soft or collinear, then the factorized formula will hold whether it is put in $\bra{X_\scs}$ or in the appropriate $\bra{X_\cci}$.
For example, we can place all the soft-collinear momenta in the soft sector by designating any particle with energy less than some $\Lambda$ as soft, and then draw cones of size $R$ around each of the hard directions to distribute particles in the collinear sectors. With such hard cutoffs, one can then square the matrix elements on the right-hand side of Eq.~\eqref{QEDmain} and perform the phase-space
integrals over the appropriate measurement function to get a differential distribution. The result will agree at leading power with the the distribution computed using the left-hand side of Eq.~\eqref{QEDmain} in the limit $R\to 0$ and $\Lambda\to 0$. 

There are two problems with the hard-cutoff prescription for resolving the soft-collinear ambiguity. The first is practical: introducing an extra scale makes the relevant calculations nearly impossible. Moreover, the cutoff dependence may not exactly cancel in the factorized expression and therefore one must either take $R\to 0$ and $\Lambda\to0$ after the calculation or live with
power corrections in these cutoffs. 
 The second is conceptual: the cutoffs violate factorization in the following sense. There will in general be leading-power dependence on the cutoff in the soft and collinear sectors separately (terms like $\frac{1}{\tau} \ln R $, for example) which only cancel when the sectors are combined. 
 Thus the two sectors are not completely separated. 

It would be great if we could simply perform phase-space integrals over each sector separately including all momenta. This is not as crazy as it sounds. We know that including very energetic virtual momenta in the soft or collinear sectors causes no problem, since the modification can always be compensated for in the matching coefficient ($\cC_\cO(\Sij)$ in Eq.~\eqref{QEDmain}). Indeed, effective theories always have different ultraviolet (UV) structure from the full theories to which they are matched. For example, in SCET, there are $\frac{1}{\varepsilon^2}$ UV poles at 1-loop in dimensional regularization, while in full QCD, one only ever has $\frac{1}{\varepsilon}$ poles. In fact, these double poles allow for the resummation of Sudakov double logarithms in SCET using the renormalization group. We also know that one does not have to distinguish soft from collinear momenta in loops when using  Eq.~\eqref{QEDmain}: the overcounting is compensated for by the vacuum matrix elements in the denominator of this 
equation. Thus, we have good reason to believe that 
subtractions similar
to the denominator factors in Eq.~\eqref{QEDmain} can be added to this formula to allow for unrestricted phase-space integrals.

Removing the overcounting of soft and collinear momenta has been addressed in the traditional approach to factorization, for certain observables~\cite{Akhoury:1998gs,Collins:1989bt,Collins:1989gx}. There, the soft limit of collinear momenta is compensated for with eikonal jet functions \cite{Berger:2003iw}. In SCET, the overcounting
can be formally avoided by not including the zero-momentum bin in any of the collinear sectors~\cite{Manohar:2006nz}. This exclusion translates
into a subtraction diagram-by-diagram. This zero-bin
subtraction is necessary in SCET because the same soft-collinear momentum region in QCD is represented by multiple fields in the effective
theory (similar overcounting is present in other effective theories, such as NRQCD). In \cite{Lee:2006nr, Idilbi:2007ff,Idilbi:2007yi} the two prescriptions were
shown to be equivalent. Alternatively, in the method-of-regions approach to SCET~\cite{Beneke:1997zp,Beneke:2002ph,Beneke:2002ni,Becher:2014oda} the overcounting is sidestepped through careful consideration of the analytic properties
of the contributions from different sectors. We briefly review these approaches and contrast them with our approach in Section~\ref{sec:compare}.

The formulation of factorization in \Tree{} and \Loop{} and Eq.~\eqref{QEDmain} is intermediate between traditional QCD and SCET. It provides a precise formulation of factorization purely in terms the fields in full QCD, but has a factorized form with a natural effective field theory interpretation.
It is based on the observation of Freedman and Luke \cite{Freedman:2011kj} that the unwieldy Feynman rules of SCET can be avoided
and the effective Lagrangian taken simply as the direct sum of $N+1$ copies of the QCD Lagrangian, corresponding to $N$ collinear sectors and a soft sector. The formulation in \Tree{} and \Loop{} can be thought of as a generalization of the Freedman-Luke proposal, equivalent but simpler at leading power, and that addresses the soft-collinear overlap of virtual momenta. In this paper, we extend the formulation so that phase-space integrations can be done without explicit cutoffs on the momenta of various sectors.

There are two main results in this paper. First, in Section~\ref{sec:AmpLevel}, we show how the specification of which sector a gluon belongs to can be removed at the amplitude level.
More precisely, suppose we have an amplitude $\cM(p_1, \ldots, p_n, q_1 \cdots q_m)$ with $n$ hard momenta and $m$ other momenta in QCD. We show how an approximation to $\cM$ which
we call $\cM_\text{sub}$ can be derived with the property that when any of the $q_i$ become soft or collinear to any of the $p_i$, $\cM_\text{sub}$ agrees with $\cM$ at leading power. 
That is, one does not have to specify which sector the $q_i$ belong to -- the matrix element is correct no matter what.
While $\cM_\text{sub}$ is not a factorized product of matrix elements, it is the sum of factorized products of matrix elements of fields and Wilson lines. Each term in this product 
is simpler than full QCD. Thus such a subtracted matrix element may be integrable analytically and therefore provide a useful basis for a subtraction scheme in fixed-order QCD. 

The second result, in Section~\ref{sec:CrossSection}, is a derivation of how at the amplitude-squared level factorization can be preserved and phase-space cutoffs removed for certain inclusive event shapes. Although the result of this section agrees with the eikonal-jet function subtraction method of traditional QCD (which is itself equivalent to SCET), we believe our derivation elucidates
some subtleties and makes the procedure more systematic. In addition, we present explicit 1-loop formulas for various relevant soft and jet functions, with and without cutoffs and with different regulators. These formulas demonstrate which objects are infrared safe, cutoff-dependent, and well-defined. Some calculational details are relegated to the appendix.
Section~\ref{sec:compare} contrasts our approach with previous approaches. We conclude in Section~\ref{sec:conclusions}.

\section{Factorization at the amplitude level}
\label{sec:AmpLevel}
We begin by quickly reviewing the notation and main results of \Tree{} and \Loop{}. These papers showed that factorization holds for massless particles whose momenta
are either soft or collinear to one of $N$ directions $n_\ccj^\mu$. States with particles of these momenta are
written as $\bra{X_\ccO\cdots X_\ccN; X_\scs}$. The hard scale (such as the center-of-mass energy) is denoted as $Q$
and scaling parameters $\lambda_\ccj$ are defined each collinear sector and $\lambda_\scs$ for the soft sector. Momenta in each collinear
sector scale as 
\begin{equation}
\bra{X_\ccj} = \bra{\ldots, q_\ccj,\ldots} 
\imply
\frac{1}{Q}(n_\ccj\cdot q_\ccj, \bar n_\ccj\cdot q_\ccj, q_\ccj^\perp ) \sim (\lambda_\ccj^2,1,\lambda_\ccj)
\end{equation}
and momenta in the soft sector scale like
\begin{equation}
\bra{X_\scs} = \bra{\ldots, k_\scs,\ldots} 
\imply
\frac{1}{Q}(n_\ccj\cdot k_\scs, \bar n_\ccj\cdot k_\scs, k_\scs^\perp ) \sim (\lambda_\scs,\lambda_\scs,\lambda_\scs)
,\quad\forall \ccj
\end{equation}
For simplicity, assume the scattering process under consideration is the decay of a heavy particle mediated by an operator $\cO$ in QED (to avoid cumbersome color indices of QCD).
Then, the factorization formula takes the form of Eq.~\eqref{QEDmain}:
\be
\bra{X_\ccO\cdots X_\ccN; X_\scs}\cO\ket{0}
 \;\LPeq\; 
\cC_\cO(\Sij) \, \frac{\bra{X_\ccO} \bar\psi W_\ccO \ket{0}}{\bra{0} Y_\ccO^\dg W_\ccO \ket{0}} \,\cdots\, \frac{\bra{X_\rN} W_\rN^\dg\psi \ket{0}}{\bra{0} W_\rN^\dg Y_\rN \ket{0}}
\;\bra{X_\scs} Y_\ccO^\dg \cdots Y_\rN \ket{0}
\nonumber
\ee
Here $\cC_\cO(\Sij)$ is a finite function of the large products of the net momentum in each jet, $\Sij = P_\cci\cdot P_\ccj$; it
does not depend on the small power-counting parameters, $\lambda_\ccj$ of $\lambda_\scs$. The Wilson lines, $W_\ccj^\dg$ and $Y_\ccj^\dg$, are defined in QCD as follows:
\be
Y_j^\dg = P \bigg\{\exp\bigg[ig\int_0^\infty ds\, n_\ccj\cdot A(x+s\,n_\ccj)\, e^{-\varepsilon s} \bigg]  \bigg\}
\ee
and
\be
W_j^\dg = P \bigg\{\exp\bigg[ig\int_0^\infty ds\, t_\ccj\cdot A(x+s\,t_\ccj)\, e^{-\varepsilon s} \bigg]  \bigg\} 
\ee
where $t_\ccj^\mu$ are some lightlike directions assumed not collinear to their associated $n_\ccj^\mu$. The $P\{\}$ denotes path ordering; in QED the path ordering is trivial and the electromagnetic charge is $e=-g$.
\Eq{QEDmain} is an equality at leading power in all of $\lambda_\ccj$ and $\lambda_\scs$ separately. For many applications, such as for thrust, one takes $\lambda^2_\ccj = \lambda_\scs$ for all $j$; in the SCET literature, this power counting is referred to as SCET$_{\mathrm I}$ \cite{Bauer:2000yr}. For recoil sensitive observables like jet broadening, one takes $\lambda_\ccj = \lambda_\scs$ as in SCET${}_{\mathrm{II}}$ \cite{Bauer:2002nz}. The factorization in Eq.~\eqref{QEDmain} holds for any relative scaling.

The important physics contained in \Eq{QEDmain} is that each factor on the right-hand side represents a different factorized sector: the Wilson coefficient, $\cC_\cO(\Sij)$, represents all of the hard physics and must be IR-insensitive. Each collinear sector is represented by the ratio $\bra{X_\ccj} W_\ccj^\dg\psi \ket{0} \big/ \bra{0} W_\ccj^\dg Y_\ccj \ket{0}$ and contains only $n_\ccj$-collinear IR divergences. Finally, the soft sector is fully described by the matrix element, $\bra{X_\scs} Y_\ccO^\dg \cdots Y_\rN \ket{0}$, which contains all of the soft divergences of the full amplitude on the left-hand side of \Eq{QEDmain}.

One attractive feature of \Eq{QEDmain} is that each matrix element is constructed out of full-theory operators and evaluated using the full-theory Lagrangian; there are no additional subtractions/prescriptions needed, just simple QCD/QED Feynman rules. Moreover, the power counting is a consequence only of the scaling of the external momenta in the states $\bra{X_\ccO\cdots X_\ccN; X_\scs}\cO\ket{0}$. An obvious fact with important repercussions is that \Eq{QEDmain} is not valid when any of the momenta in a given sector does not obey the scaling that is associated with that sector. Consequently, one cannot, for example, integrate over the entire phase space of one of the external momenta in \Eq{QEDmain} because it would enter the scaling regime of other sectors.

Therefore, when calculating cross sections by squaring \Eq{QEDmain} one can either integrate over the phase space $d\Pi_{X_\ccj}$ with cutoffs in the integrals restricting each integral to be within the collinear region, or one  can try to extend the integrations to the entire phase space and perform a subtraction that gets rid of the errors that we introduced by extending $d\Pi_{X_\ccj}$ to the entire phase space. Introducing cutoffs to integrals is incredibly tedious and produces new scales in the effective theory that obscure factorization (as shown explicitly in Section~\ref{sec:CrossSection}). The subtraction procedure is the only reasonable way forward. We next discuss subtractions at the amplitude level, and discuss subtractions at the cross section level in Section~\ref{sec:CrossSection}.

\subsection{Example subtractions}
Consider the case of a $q \bar{q} g$ final state, with quark momenta $p_\ccO^\mu$ and $p_\ccT^\mu$ in different directions and the gluon momentum $q^\mu$. Suppose we want to integrate over the gluon momenta inclusively. We can do so using Eq.~\eqref{QEDmain}
if when $q \parallel p_\ccO$ we use 
\begin{equation}
\cM_\ccO(p_\ccO,p_\ccT,q) \equiv
\frac{\bra{p_\ccO;q} \bar\psi W_\ccO \ket{0}}{\bra{0} Y_\ccO^\dg W_\ccO \ket{0}}
 \frac{\bra{p_\ccT} W_\ccT^\dg\psi \ket{0}}{\bra{0} W_\ccT^\dg Y_\ccT  \ket{0}}
 \bra{0} Y_\ccO^\dg Y_\ccT \ket{0} \; ,
 \end{equation}
if $q \parallel p_\ccT$,  we use
\begin{equation}
\cM_\ccT(p_\ccO,p_\ccT,q)  \equiv 
\frac{\bra{p_\ccO} \bar\psi W_\ccO \ket{0}}{\bra{0} Y_\ccO^\dg W_\ccO \ket{0}}
 \frac{\bra{p_\ccT;q} W_\ccT^\dg\psi \ket{0}}{\bra{0} W_\ccT^\dg Y_\ccT  \ket{0}}
 \bra{0} Y_\ccO^\dg Y_\ccT \ket{0} \; ,
 \end{equation}
and if $q$ is soft, we use
\begin{equation}
\cM_\scs(p_\ccO,p_\ccT,q)  \equiv 
\frac{\bra{p_\ccO} \bar\psi W_\ccO \ket{0}}{\bra{0} Y_\ccO^\dg W_\ccO \ket{0}}
 \frac{\bra{p_\ccT} W_\ccT^\dg\psi \ket{0}}{\bra{0} W_\ccT^\dg Y_\ccT  \ket{0}}
 \bra{q} Y_\ccO^\dg Y_\ccT \ket{0}
 \end{equation}
However we split up the integration regions (say with a soft energy cutoff $\Lambda$ and cone radius $R$) the dependence on the split (on $\Lambda$ and $R$) will drop out at leading power when all three contributions are added. Nevertheless, it would be nice to have an expression that we could simply integrate over $q$ without ever introducing $\Lambda$ and $R$ in the first place.

To proceed, we first examine the consequences of soft-collinear factorization for the operator $\cO_{\bar\psi W} = \bar{\psi} W_\ccO$ (rather than a local QCD operator like $\bar\psi \psi$). The all-orders proof of factorization in \Loop{} applies to $\cO_{\bar\psi W}$. In particular, if we have a state with momenta $p_1\cdots p_n$ all of which are collinear to each other 
as well as momenta $q_1 \cdots q_m$ all of which are soft, then
\be
\bra{p_1\cdots p_n ; q_1 \cdots q_m} \bar\psi W_\ccO \ket{0} \LPeq
 C_{\bar\psi W} \bra{ p_1 \cdots p_n} \bar\psi W_\ccO \ket{0}\frac{\bra{q_1 \cdots q_m} Y_\ccO^\dag W_\ccO \ket{0}}{\bra{0} Y_\ccO^\dg W_\ccO \ket{0}}
\label{genstate}
\ee
for some $C_{\bar\psi W}$.
 To determine $C_{\bar\psi W}$, we note that $C_{\bar\psi W}$ does not depend on how the momentum in the collinear and soft sectors are distributed; this equation holds for any $n>0$ and any $m\ge 0$. In particular, if we take $m=0$ then the two sides are identical (and agree at leading power) if and only if $C_{\bar\psi W}=1$. Thus we must have $C_{\bar\psi W}=1$ for any states.

As a special case, Eq.~\eqref{genstate} implies that for one collinear and one soft momentum
\be
\frac{
\bra{p_\ccO;q} \bar\psi W_\ccO \ket{0}}{\bra{0} Y_\ccO^\dg W_\ccO \ket{0}}
~~
\overset{q~\text{soft}}{
\LPeq}
~~
 \frac{\bra{p_\ccO} \bar\psi W_\ccO \ket{0}}{\bra{0} Y_\ccO^\dg W_\ccO \ket{0}}
 \frac{\bra{q} Y_\ccO^\dag W_\ccO \ket{0}}{\bra{0} Y_\ccO^\dg W_\ccO \ket{0}}
\label{softlim}
\ee
Similarly, applying the general factorization formula to $\cO =  Y_\ccO^\dg Y_\ccT$, we get
\be
 \bra{q} Y_\ccO^\dg Y_\ccT \ket{0} 
~
\overset{\;\,q~ \parallel~ p_\ccO}{
\LPeq}
~~
\frac{\bra{q} Y_\ccO^\dg W_\ccO \ket{0}}{\bra{0} Y_\ccO^\dg W_\ccO \ket{0}}  
	\times \bra{0} Y_\ccO^\dg Y_\ccT \ket{0} 
\label{colllim}
\ee
In this case one can see that the Wilson coefficient is $1$ to all orders by using the proof in \Loop{} that the factorization theorem 
is independent of the collinear Wilson-line direction, $t_\ccO$, and then choosing $t^\mu_\ccO = n^\mu_\ccT$, so that $W_\ccO = Y_\ccT$.

With these results, we can now analyze the following all-loop-order subtracted matrix element:
\pagebreak
\begin{multline}
\cM_\text{sub}(p_\ccO,p_\ccT,q) \equiv 
\left\{\frac{\bra{p_\ccO;q} \bar\psi W_\ccO \ket{0}}{\bra{0} Y_\ccO^\dg W_\ccO \ket{0}}
-
\frac{\bra{p_\ccO} \bar\psi W_\ccO \ket{0}}{\bra{0} Y_\ccO^\dg W_\ccO \ket{0}}
\frac{\bra{q} Y_\ccO^\dag W_\ccO \ket{0}}{\bra{0} Y_\ccO^\dg W_\ccO \ket{0}}
\right\}
 \frac{\bra{p_\ccT} W_\ccT^\dg\psi \ket{0}}{\bra{0} W_\ccT^\dg Y_\ccT  \ket{0}}
 \bra{0} Y_\ccO^\dg Y_\ccT \ket{0}
\\
+
\frac{\bra{p_\ccO} \bar\psi W_\ccO \ket{0}}{\bra{0} Y_\ccO^\dg W_\ccO \ket{0}}
\left\{\frac{\bra{p_\ccT;q} \bar\psi W_\ccT \ket{0}}{\bra{0} Y_\ccT^\dg W_\ccT \ket{0}}
-
\frac{\bra{p_\ccT} \bar\psi W_\ccT \ket{0}}{\bra{0} Y_\ccT^\dg W_\ccT \ket{0}}
\frac{\bra{q} Y_\ccT^\dag W_\ccT \ket{0}}{\bra{0} Y_\ccT^\dg W_\ccT \ket{0}}
\right\}
 \bra{0} Y_\ccO^\dg Y_\ccT \ket{0}
\\
+
\frac{\bra{p_\ccO} \bar\psi W_\ccO \ket{0}}{\bra{0} Y_\ccO^\dg W_\ccO \ket{0}}
 \frac{\bra{p_\ccT} W_\ccT^\dg\psi \ket{0}}{\bra{0} W_\ccT^\dg Y_\ccT  \ket{0}}
 \bra{q} Y_\ccO^\dg Y_\ccT \ket{0}
\label{sub123}
\end{multline}
If we take $q$ soft, then neither of the first two lines contribute by Eq.~\eqref{softlim}, and the result is given by the third line which is the correct
leading power matrix element $\cM_\scs$. When
$q \parallel p_\ccO$, then neither term in the second line is IR sensitive and the subtraction term in the first line (which is IR-sensitive)
is canceled by collinear limit of the third line, using Eq.~\eqref{colllim}. Thus, only the first term on the first line contributes at leading power in this limit, in agreement with $\cM_\ccO$. The analogous argument works for the $q \parallel p_\ccT$ limit. We conclude that $\cM_\text{sub}(p_\ccO,p_\ccT,q)$ agrees 
with full QCD at leading power for any $q$. Thus, we can integrate $\cM_\text{sub}$ over phase space without splitting the soft and collinear sectors.

To be explicit, we can evaluate Eq.~\eqref{sub123} in perturbation theory. At tree-level,
\begin{multline}
\cM(p_\ccO,p_\ccT,q) \overset{\text{tree}}= \bar u(p_1) \bigg\{
\Big(\frac{-g\Sl\epsilon_q (\Sl p_1 + \Sl q)}{2p_1\cdot q} 
	+ \frac{g t_1\cdot \epsilon_q}{t_1\cdot q} \Big)
- 
\Big(\frac{-g n_1\cdot \epsilon_q}{n_1\cdot q}
	+ \frac{g t_1\cdot \epsilon_q}{t_1\cdot q} \Big)
\bigg\} v(p_2)
\\+ 
\bar u(p_1) \bigg\{
\Big( \frac{g(\Sl p_2 + \Sl q)\Sl\epsilon_q}{2p_2\cdot q} 
	+ \frac{-g t_2\cdot \epsilon_q}{t_2\cdot q}\Big)
- 
\Big(\frac{-g t_2\cdot \epsilon_q}{t_2\cdot q}
	+ \frac{g n_2\cdot \epsilon_q}{n_2\cdot q} \Big)
\bigg\} v(p_2)
\\+
\bar u(p_1) \Big( 
\frac{-g n_1\cdot \epsilon_q}{n_1\cdot q}
	+ \frac{g n_2\cdot \epsilon_q}{n_2\cdot q}
\Big)v(p_2)
\label{mform}
\end{multline}
where each term in round brackets corresponds to one of the matrix elements containing the gluon, and thereby each satisfies the Ward identity separately. From the explicit expression in \Eq{mform} it is easy to check that each soft and collinear limit works out exactly as stated in the paragraph after \Eq{sub123}. It can also be seen that the $t_j$ dependent terms cancel out completely as do the soft terms containing $n_j$ at this order, leaving:
\be
\cM_\text{sub}(p_\ccO,p_\ccT,q) \overset{\text{tree}}= \bar u(p_1) \bigg(
\frac{-g\Sl\epsilon_q (\Sl p_1 + \Sl q)}{2p_1\cdot q} 
+ \frac{g(\Sl p_2 + \Sl q)\Sl\epsilon_q}{2p_2\cdot q} 
\bigg) v(p_2)
\overset{\text{tree}}=
\cM(p_\ccO,p_\ccT,q) 
\ee
So the the full matrix element of QED is reproduced exactly in this case. Of course, for more complex calculations we expect $\cM$ to only reproduce the full-theory matrix element at leading power, rather than be exactly equal to it.

\subsection{General amplitude-level subtraction}
The generalization of Eq.~\eqref{sub123} for arbitrary collinear and soft sectors is
\begin{multline}
\label{SubTry2}
\bra{X_\ccO\cdots X_\ccN ; X_\scs;q}\cO\ket{0}
 \LPFeq 
 \frac{\bra{X_\ccO} \bar\psi W_\ccO \ket{0}}{\bra{0} Y_\ccO^\dg W_\ccO \ket{0}}
\cdots
\frac{\bra{X_{\ccN}} W^\dg_{\ccN}\psi \ket{0}}{\bra{0} W_{\ccN}^\dg Y_{\ccN} \ket{0}}
 \bra{X_\scs , q} Y_\ccO^\dg \cdots Y_\ccN \ket{0} 
\\
+
\sum_{\cci=1}^N
\frac{\bra{X_\ccO} \bar\psi W_\ccO \ket{0}}{\bra{0} Y_\ccO^\dg W_\ccO \ket{0}}
\cdots\bigg\{  \frac{\bra{X_\cci, q} W_\cci^\dg\psi \ket{0}}{\bra{0} W_\cci^\dg Y_\cci \ket{0}} \bigg\}_{\SoftSub{q  }}
\cdots
\frac{\bra{X_{\ccN}} W^\dg_{\ccN}\psi \ket{0}}{\bra{0} W_{\ccN}^\dg Y_{\ccN} \ket{0}}
\bra{X_\scs} Y_\ccO^\dg \cdots Y_\ccN \ket{0}
\end{multline}
where the $\{\}\SoftSubNO{q} $ notation means the operator matrix element corresponding to having subtracted the $q\to$ soft  limit. 
To be explicit, we can use the notation $\SL(q)$ as in \Loop{} for the leading order contribution in the $q\to $ soft limit. 
Then
\begin{align}
\bigg\{  \frac{\bra{X_\cci, q} W_\cci^\dg\psi \ket{0}}{\bra{0} W_\cci^\dg Y_\cci \ket{0}} \bigg\}_{\SoftSub{q}}  
&\equiv
 \frac{\bra{X_\cci, q} W_\cci^\dg\psi \ket{0}}{\bra{0} W_\cci^\dg Y_\cci \ket{0}} 
-
\bigg(\frac{\bra{X_\cci, q} W_\cci^\dg\psi \ket{0}}{\bra{0} W_\cci^\dg Y_\cci \ket{0}} \bigg)_{\SL(q)}
\notag\\&=
 \frac{\bra{X_\cci, q} W_\cci^\dg\psi \ket{0}}{\bra{0} W_\cci^\dg Y_\cci \ket{0}} 
-
 \frac{\bra{X_\cci} W_\cci^\dg\psi \ket{0} \bra{q} W_\cci^\dg Y_\cci \ket{0}}{\bra{0} W_\cci^\dg Y_\cci \ket{0}^2} 
\end{align}
This subtracted quantity is exactly the same as what was used in \Eq{sub123} and vanishes at leading power in the $q\to $ soft limit by \Eq{softlim}. \Eq{SubTry2} is a sum of factorized expressions which
agrees at leading power with full QCD in any soft or collinear limit of $q$.

To generalize to multiple gluons or quarks with momenta $q_i$, the analogous formula is easiest to define recursively. For example,
adding a second gluon to \Eq{SubTry2}, we can either place it in the soft matrix element, or in a collinear matrix element. If it
is in the collinear matrix element, we must subtract off the soft limit. Thus we get a sum of terms:
\begin{multline}
\label{SubTry3} 
\bra{X_\ccO \cdots X_\ccN; X_\scs; q_1, q_2}\cO\ket{0}
 \LPFeq
 \frac{\bra{X_\ccO} \bar\psi W_\ccO \ket{0}}{\bra{0} Y_\ccO^\dg W_\ccO \ket{0}}
\cdots
\frac{\bra{X_{\ccN}} W^\dg_{\ccN}\psi \ket{0}}{\bra{0} W_{\ccN}^\dg Y_{\ccN} \ket{0}}
 \bra{X_\scs , q_1,q_2} Y_\ccO^\dg \cdots Y_\ccN \ket{0} 
\\ +
 \sum_{\cci=1}^N
\cdots \bigg\{ \frac{\bra{X_\cci, q_1} W_\cci^\dg\psi \ket{0}}{\bra{0} W_\cci^\dg Y_\cci \ket{0}} \bigg\}_{\SoftSub{q_1}}  
\cdots
\bra{X_\scs,q_2} Y_\ccO^\dg \cdots Y_\ccN \ket{0}
\\ +
 \sum_{\cci=1}^N
\cdots \bigg\{ \frac{\bra{X_\cci, q_2} W_\cci^\dg\psi \ket{0}}{\bra{0} W_\cci^\dg Y_\cci \ket{0}} \bigg\}_{\SoftSub{q_2}}  
\cdots
\bra{X_\scs,q_1} Y_\ccO^\dg \cdots Y_\ccN \ket{0}
\\ +
 \sum_{\cci,\ccj=1}^N
\cdots \bigg\{\frac{\bra{X_\cci, q_1} W_\cci^\dg\psi \ket{0}}{\bra{0} W_\cci^\dg Y_\cci \ket{0}} \bigg\}_{\SoftSub{q_1}}  
\cdots \bigg\{\frac{\bra{X_\ccj, q_2} W_\ccj^\dg\psi \ket{0}}{\bra{0} W_\ccj^\dg Y_\ccj \ket{0}} \bigg\}_{\SoftSub{q_2}} 
\cdots
\bra{X_\scs} Y_\ccO^\dg \cdots Y_\ccN \ket{0}
\end{multline}
where the $\cdots$ represent the other collinear matrix elements which do not contain any $q$'s. In the last line,
when $\cci=\ccj$ the soft subtraction must be done iteratively to ensure that the subtraction mitigates the soft enhancement in any order of limits of $q_1$ and $q_2$ going soft. That is,
\begin{multline}
 \frac{\bra{X_\ccj, q_1,q_2} W_\ccj^\dg\psi \ket{0}}{\bra{0} W_\ccj^\dg Y_\ccj \ket{0}} \bigg|_{\SoftSub{q_1,q_2}} 
 \equiv
 \frac{\bra{X_\ccj, q_1,q_2} W_\ccj^\dg\psi \ket{0}}{\bra{0} W_\ccj^\dg Y_\ccj \ket{0}} 
\;-\; \bigg(
	\frac{\bra{X_\ccj, q_1,q_2} W_\ccj^\dg\psi \ket{0}}{\bra{0} W_\ccj^\dg Y_\ccj \ket{0}}  
	\bigg)_{\SL(q_1,q_2)}
\\ \label{SubTry4}
\qquad-\; \Bigg(
	\frac{\bra{X_\ccj, q_1,q_2} W_\ccj^\dg\psi \ket{0}}{\bra{0} W_\ccj^\dg Y_\ccj \ket{0}}  
\;-\; \bigg(
	\frac{\bra{X_\ccj, q_1,q_2} W_\ccj^\dg\psi \ket{0}}{\bra{0} W_\ccj^\dg Y_\ccj \ket{0}}  	
	\bigg)_{\SL(q_1,q_2)} \Bigg)_{\SL(q_1)}
\\
\qquad-\; \Bigg(
	\frac{\bra{X_\ccj, q_1,q_2} W_\ccj^\dg\psi \ket{0}}{\bra{0} W_\ccj^\dg Y_\ccj \ket{0}}  
\;-\; \bigg(
	\frac{\bra{X_\ccj, q_1,q_2} W_\ccj^\dg\psi \ket{0}}{\bra{0} W_\ccj^\dg Y_\ccj \ket{0}}  	
	\bigg)_{\SL(q_1,q_2)} \Bigg)_{\SL(q_2)}
\end{multline}
where $\SL(q_1,q_2)$ means taking the leading-power expression in the $q_1,q_2\to $ soft limit simultaneously and, therefore, does not drop $q_1$ with respect to $q_2$ or vice-versa. Note that, as always, we can write the soft limits in terms of amplitudes with Wilson lines using the factorization theorem of \Eq{QEDmain}. For example,
\be
\bigg(
	\frac{\bra{X_\ccj, q_1,q_2} W_\ccj^\dg\psi \ket{0}}{\bra{0} W_\ccj^\dg Y_\ccj \ket{0}}  
	\bigg)_{\SL(q_1,q_2)}
\;=\;
\frac{\bra{X_\ccj} W_\ccj^\dg\psi \ket{0}}{\bra{0} W_\ccj^\dg Y_\ccj \ket{0}}  
\frac{\bra{q_1,q_2} W_\ccj^\dg Y_\ccj \ket{0}}{\bra{0} W_\ccj^\dg Y_\ccj \ket{0}}
\ee
where we know that the Wilson coefficient will always be $1$ to all orders by the argument given after \Eq{genstate}.

That these subtractions will always work follows using the arguments of \Loop. In particular, the ``coloring algorithm'' in Section 6 of that paper is exactly the recursive soft subtraction procedure indicated by  Eqs. \eqref{sub123}, \eqref{SubTry2}--\eqref{SubTry4}. As with the algorithm in \Loop, the soft limit of any subset of the $q$'s in the $\big\{\big\}\SoftSubNO{q_1\cdots q_m}$ matrix elements are power suppressed, and they should correspondingly be colored blue. With this knowledge, it is easy to check that \Eq{SubTry3} agrees in the IR: when $q_1, q_2 \to$ soft, all of the $\big\{\big\}\SoftSubNO{q_\cci}$ matrix elements are power suppressed and only the top line survives, which gives the correct answer. When $q_1 \parallel p_\ccj$ and $q_2 \to$ soft, say, the bottom two lines are power suppressed and the $\SL(q_1)$-subtracted term cancels with the top line, leaving only the one term that matches the full-factorized formula in this limit. Similarly, all other limits can be simply checked.  The 
pattern of subtractions with more than two gluons follows exactly as with the coloring algorithm stated in generality in \Loop.

The procedure outlined in this section produces amplitudes which can be computed as a sum of factorized terms. These amplitudes, which are a new result, reproduce all of the leading-power IR-sensitive limits of the full-QCD amplitudes, at all-loop order. Each factor in each term in the sum involves matrix elements of fields and Wilson lines that are universal and simpler than the factors in the full QCD amplitude. Given these properties, an interesting application of the matrix elements derived in this section might be towards subtraction procedures for QCD calculations at NNLO or beyond. One application of subtraction methods is to split an amplitude into a universal IR-sensitive piece that is simple enough to integrate analytically and a piece that is IR-finite which could be integrated numerically \cite{Alioli:2013hba,Frixione:1995ms,Kunszt:1992tn,Catani:1996vz,GehrmannDeRidder:2005cm}. The amplitudes presented in this section could be a candidate for such a procedure at any order in perturbation theory 
and for any 
number of external 
particles.

\section{Factorization for distributions}
\label{sec:CrossSection}

Despite having many strengths, amplitudes as in \Eq{SubTry2}, are no longer factorized:  they cannot be written as a single product of terms with the same external states (in this case the collinear sectors and the soft sector are tangled). When the amplitudes are squared, the interference effects between various terms in the sum contribute at leading power, so they must all be included. Thus, while one {\it can} integrate over the momenta $q_j$ without overcounting the infrared-sensitive region, the separation between soft and collinear contributions
is no longer manifest. Moreover, it is not clear how the large logarithms associated with the leading-power IR sensitivity can be resummed using such amplitudes.

Fortunately, for certain observables, one can perform subtractions differently so that factorization is preserved at the cross-section level. 
In this section, we discuss a class of factorizing observables. Namely, we discuss observables whose measurement function, that is, the mapping from the final-state momenta to the observable, is linear in the soft and collinear momenta. These observables include many $e^+e^-$ event shapes, such as 
thrust \cite{Farhi:1977sg,Catani:1992ua,Schwartz:2007ib,Becher:2008cf,Abbate:2010xh}, 
angularities~\cite{Berger:2003iw,Berger:2003pk,Hornig:2009vb}
heavy jet mass \cite{Chien:2010kc}, the $C$ parameter \cite{Parisi:1978eg,Donoghue:1979vi,Hoang:2015hka} and jet broadening \cite{Rakow:1981qn,Ellis:1986ig,Catani:1992jc,Chiu:2011qc,Becher:2011pf}. Many hadron collider observables are also in this class~\cite{Laenen:1998qw},
 such Drell-Yan near threshold \cite{Becher:2007ty}, deep inelastic scattering as $x\to 1$ \cite{Manohar:2003vb}, direct photon production \cite{Becher:2009th,Kidonakis:2003bh}, $W/Z$ + jet~\cite{Gonsalves:2005ng, Becher:2011fc,Becher:2012xr}, jet mass \cite{Chien:2012ur,Dasgupta:2012hg}, or $t\bar{t}$ production near the hadronic threshold~\cite{Ahrens:2010zv,Kidonakis:2012rm} as well as $N$-(sub)jettiness \cite{Stewart:2010tn,Thaler:2010tr,Feige:2012vc}.

Factorization at the cross-section level for observables in this class has been understood already by traditional QCD and by Soft-Collinear Effective Theory (see above references).
The overcounting of soft and collinear integration regions is also well-understood in both approaches, and the two approaches have already been shown to be equivalent \cite{Lee:2006nr, Idilbi:2007ff,Idilbi:2007yi}. 
Unfortunately, it is challenging to extract from the literature which aspects of the removal of overcounting have been rigorously proven (in either approach) and which aspects are simply assumed. Moreover, the overcounting in phase-space integrals has not been addressed at all in the effective field theory formulation with full-theory fields \cite{Freedman:2011kj}, \Tree{},\Loop{}.
The goal of this section is to give a self-contained proof that the overcounting induced by removing phase-space cutoffs can be completely compensated for. We thereby demonstrate a form of factorization that
holds exactly at leading power at the cross-section level with no phase space cutoffs. 

\subsection{Factorization for thrust}
\label{sec:CrossSectionThrust}

For concreteness and simplicity, we begin our discussion with thrust, the paradigmatic observable whose distribution factorizes. Thrust, $T$, is defined as~\cite{Farhi:1977sg}
\be
T \equiv \frac{\sum_j | \vec{p}_j \cdot \vec{n} |} {\sum_j |\vec{p}_j|}
\ee
where $\vec{n}$ is the thrust axis, defined to maximize $T$. The region where factorization holds is where $\tau = 1-T \ll 1$. Then
\be
\tau \LPeq 1-\sum_j \frac{1}{Q}| \vec{p}_j \cdot \vec{n}|  = \frac{1}{2Q} \sum_j \Omega_\tau(p_j) 
\ee
where $Q$ is the center of mass energy and $\Omega_\tau(p)$ is the measurement function for thrust:
\be
\Omega_\tau(p) = p^- \theta(p^+ - p^-) + p^+\theta(p^- - p^+)
\label{Omegadef}
\ee
where  $p^+ = n\cdot p$ and $p^- = \bar{n} \cdot p$. 

Note that $\tau$ has the property that it is linear in the momenta: each particle momentum contributes additively to thrust, independent of the other momenta in the final state $\bra{X}$. In particular, if we decompose $\bra{X}$ into soft, collinear and hard momenta, then we can compute the contribution to thrust from each sector separately and just add the results. In other words, linearity implies
\be
\delta\Big(\tau -\frac{1}{2Q}\Omega_\tau(p_X)\Big)
=
\delta\Big(\tau-\frac{1}{2Q} p_\scs -\frac{1}{2Q} p_\ccO -\frac{1}{2Q} p_\ccT 
-\frac{1}{2Q} p_\hch 
\Big) 
\label{delta}
\ee
where $p_\scs$ is the sum of $\Omega_\tau(k)$ over the soft momenta, $p_\ccO$ and $p_\ccT$ the sum over collinear momenta in each direction and $p_h$ the sum over the remaining
momenta. Writing the argument of the $\delta$-function as a sum lets us turn products of matrix elements into convolutions.

\begin{figure}[t!]
	\begin{center}
\includegraphics[scale=.8]{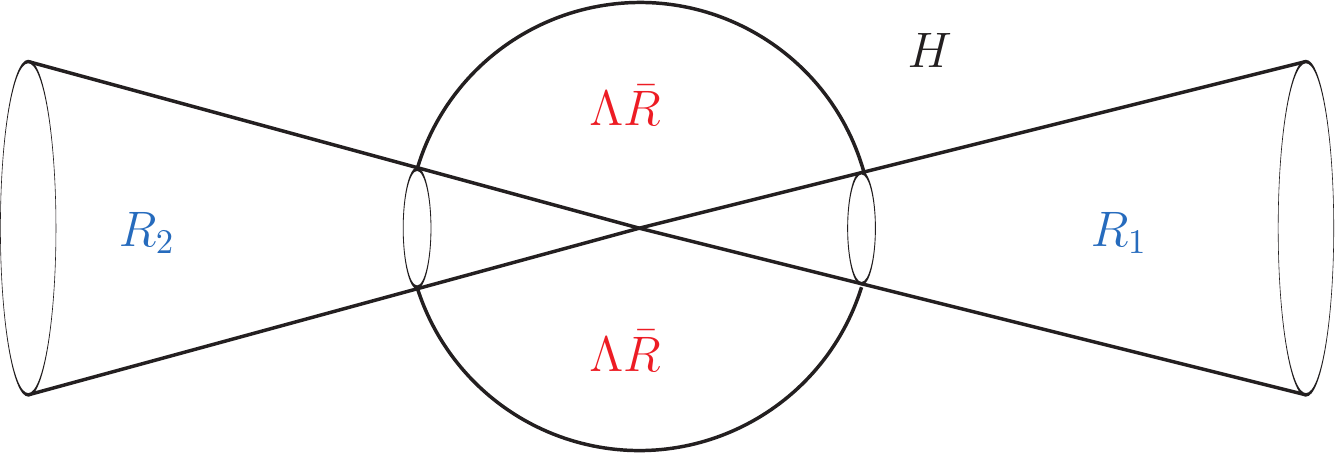}
\caption{Mutually exclusive sectioning of phase space into $\ccj$-collinear, soft and hard momentum labelled by $\ccR_\ccj$, $\scL\scRb$ and $H$, respectively.
In explicit formulas in this paper, $R$ is treated as a rapidity variable: $R=\tan^2\frac{\theta}{2}$, with $\theta$ the opening angle of the cone.}
\label{fig:sectors}
	\end{center}
\end{figure}

To be concrete let us place the momenta into sectors using hard cuts: we draw cones of angular size $R$  around the $\vec{n}$ and $-\vec{n}$ collinear directions and a ball of size $\Lambda$ 
around the origin; anything in the cones but not the ball is collinear, $\bra{X_\ccj}$, anything in the ball but not the cones is soft, $\bra{X_\scs}$. For later convenience, we include the soft-collinear radiation, which is in both the ball and the a cone, in the collinear sector (we could equally well have put it in the soft sector). 
Anything not in the cone or ball is called hard, $\bra{X_H}$. 
This breakdown of phase space is shown in Fig~\ref{fig:sectors}. 

Consider the thrust distribution in full QCD mediated by the operator $\cO = \bar{\psi} \gamma^\mu \psi$. That is,
\be
\frac{d\sigma}{d \tau}
= \sum_{X} \int d \Pi_X |\bra{X}\cO\ket{0}|^2 \delta\Big(\tau -\frac{1}{2Q}\Omega_\tau(p_X) \Big)
\ee
where the sum is over all possible final states $\bra{X}$ and the normalization and momentum-conserving $\delta$-function are
left implicit.

When $\tau$ is small,
only states of the form $\bra{X} = \bra{X_\ccO}\bra{X_\ccT}\bra{X_\scs}$ contribute at leading power in $\tau$.
With the hard phase-space cuts in place, the factorization formula at the amplitude level, \Eq{QEDmain}
along with Eq.~\eqref{delta},
 immediately generates a factorization formula for the thrust distribution:
\be
\frac{d\sigma}{d \tau} \LPeq H \times  S^{\scL\scRb} \conv J^{\ccR_\ccO} \conv J^{\ccR_\ccT}
\label{factcut}
\ee
Here $H = |\cC|^2$ refers to the hard function (the square of the Wilson coefficient in the factorization formula). $J^{\ccR_\ccO}$ and $J^{\ccR_\ccT}$ are jet functions with restricted phase-space integrals:
\be
J^{\ccR_\ccO}(\tau) = \sum_{X_\ccO} \int d \Pi_{X_\ccO} 
\left| \frac{\bra{X_\ccO} \bar\psi W_\ccO \ket{0}}{\bra{0} Y_\ccO^\dg W_\ccO \ket{0}} \right|^2
\delta\big(\tau-p^+_{X_\ccO} \big)
\label{jetcutdef}
\ee
with $X_\ccO$ the set of states all of whose momenta are within an angular distance $R$ of the $n^\mu$ direction.  Note that we have modified the measurement function from $\Omega_\tau(p)$ in Eq.~\eqref{Omegadef} to simply $p^+$. This is allowed since all the momenta
in the cone necessarily have $p^- > p^+$ so the step functions in Eq.~\eqref{Omegadef} can be evaluated explicitly. Analogously, the jet function, $J^{\ccR_\ccT}$, will have the measurement function replaced by $p^-$. 
Lastly, $S^{\scL\scRb}$ is the phase-space restricted soft function
\be
S^{\scL \scRb}(\tau) = \sum_{ X_{\scs}
} \int d \Pi_{X_{\scs}} 
\left| \bra{X_{\scs}} Y_\ccO^\dg \cdots Y_\rN \ket{0} \right|^2
\delta \Big(\tau- \frac{1}{2Q}\Omega_\tau(p_{X_{\scs}})\Big)
\label{softres}
\ee
Here the states have momenta which are not collinear, that is, they are an angular distance greater than $R$ from all jets, and they have energy less than $\Lambda$.

The equivalence in Eq.~\eqref{factcut} holds at leading power in $\tau$ only if $R$ and $\Lambda$ are small enough so that the
collinear radiation is collinear and the soft radiation is soft. More precisely, it holds at leading power in $R$ and $\Lambda$,
meaning that the two sides my differ by terms of order $R$ or order $\Lambda$ which vanish as $R\to0$ and $\Lambda\to0$. Since
the operators entering the soft and jet function are different, we do not expect the $R$ dependence to cancel exactly between them;
the factorization theorem only guarantees that it vanish at leading power.

We suspect it may be pedagogically useful to examine explicit expressions for $S^{\scL\scRb}$ and $J^\ccR$. To distinguish UV divergences from IR divergences we include an off-shellness regulator $\omega$ for the IR (see \Eq{omegareg} in the Appendix) and analytically continue to $d=4-2\e$ dimensions for the UV. Some intermediate steps are relegated to Section~\ref{sec:jets} in the Appendix. For the unrenormalized jet function at finite $R$, we find
\begin{align}
J^{\ccR_\ccj}(\tau)  &\LPeq \delta(\tau)+ \frac{\alpha_{s} C_{F}}{ 2 \pi } \left(\frac{ \mu^{2} }{ Q^{2} }\right)^{\e} 
 \bigg\{ \delta(\tau)  \left(\frac{2}{\e^2} + \frac{3}{2 \e}   + \frac{7}{2} + \frac{\pi^2}{6} \right)   
\notag\\&
+ \delta(\tau) \left( -\frac{2}{\e} \ln{\omega} -2 \ln{\omega} \ln R +  2\ln^2{\omega}  + \cO(R) \right)
- \left( \frac{3}{2}  -2 \ln R \right) \left[\frac{1}{\tau} \right]_{+} - 2 \left[ \frac{\ln \tau}{\tau}\right]_+ \bigg\}
\label{jetcutExpResult}
\end{align}
This expression includes both the real and virtual contributions to $\bra{X_\ccO} \bar\psi W_\ccO \ket{0}$ and the purely virtual contributions to $\bra{0} Y_\ccO^\dg W_\ccO \ket{0}$ in \Eq{jetcutdef}. Note that it has
$\frac{1}{\e^2}$ UV poles, which come from the virtual graphs. It also has an overlapping UV-IR singularity (the $\frac{1}{\e}\ln\omega$ term on the second line). This singularity, which cannot be removed through local counterterms, comes from loops involving the Wilson lines which go to infinite energy collinear to one of the Wilson line directions. 
The $\log R$ dependence in \Eq{jetcutExpResult} comes from the soft-collinear region of the restricted phase-space integral. Indeed, it cannot come from the 
collinear-but-not soft region, since at arbitrarily small $\tau$, the radiation is forced arbitrarily close to the jet axis and must be a finite distance from the cone boundary.
In the soft-collinear region, the radiation can be soft but an angular distance $R$ from the axis, so there can be $R$ dependence at leading power in $\tau$. 
That the $\ln R$ dependence comes from only the soft-collinear region is to be expected if it is to be completely canceled by the soft function.

The soft function outside the cones with finite $\Lambda$ is
\begin{multline}
S^{ \scL \scRb  } (\tau)  =  \delta(\tau) + C_F \frac{ \alpha_s }{ \pi}  \left( \frac{\mu^2}{ Q^2} \right)^\e \Bigg\{  
 \delta(\tau) \left( -\frac{1}{\e^2} - \frac{ 7 \pi^2}{12}  + \frac{2}{\e} \ln \omega +2 \ln \omega \ln R -2  \ln^2 \omega
+\cO(R) \right) 
\\
 -  \bigg[\frac{2 }{\tau}  \ln R  \bigg]_ +  \Bigg\}  \theta \Big(\Lambda - \frac{\tau}{R} \Big)
\\ 
+ C_F \frac{ \alpha_s }{ \pi}  \left( \frac{\mu^2}{ Q^2} \right)^\e \Bigg\{   \delta(\tau) \left[ - \frac{1}{\e^2} - \frac{ \pi^2}{4}  - 2 \left(-\frac{1}{\e} \ln \omega + \ln \omega \ln \frac{\Lambda}{Q} + \frac{1}{2} \ln^2 \omega \right) \right]
 -  \bigg[\frac{2 }{\tau }  \ln \frac{\tau Q}{\Lambda} \bigg]_+   \Bigg\} \theta\Big( \frac{\tau}{R} - \Lambda \Big)   
\label{SRL}
\end{multline} 
This function also has an incomplete cancellation between the real and virtual contributions. In particular, the virtual includes the soft-collinear region which is excluded from the real emission. Note that the $\Lambda$ dependence is entirely subleading power in $\tau$: for small $\tau$, the $\theta$ function in the third line in \Eq{SRL} vanishes 
and the other $\theta$ function evaluates to unity. 
The $R$ dependence is not subleading power as $\tau\to0$. There are also $\cO(R)$ terms not shown here but written
out in Eq.~\eqref{SRbar}.
We will come back to the cancellation of the $R$ dependence among the two jet functions and the soft function shortly.

Convolving \Eq{jetcutExpResult} for each jet with \Eq{SRL} we get
\begin{multline}
S^{\scL\scRb} \conv J^{\ccR_\ccO} \conv J^{\ccR_\ccT} \\
\LPeq  \delta(\tau) + C_F  \frac{\alpha_s}{ \pi } \left( \frac{\mu^2}{ Q^2} \right)^\e \bigg\{ \delta(\tau) \left( \frac{1}{\e^2} +  \frac{3}{2 \e}  + \frac{7}{2} - \frac{5 \pi^2}{12}  + \cO(R)  \right) - \frac{3}{2} \left[ \frac{1}{\tau} \right]_+  - 2 \left[ \frac{\ln \tau}{\tau}  \right]_{+} \bigg\}
\label{RcancelsExpResult}
\end{multline}
Note that the $\Lambda$ dependence has dropped out completely, and the $R$ dependence which is singular as $R\to 0$ has also dropped out.
It is not hard to verify that this result agrees with the full-theory result for thrust at leading power, up to the coefficient of $\delta(\tau)$ which is corrected by the hard function.

While the factorization formula for thrust in Eq.~\eqref{factcut} works, it has numerous flaws.
 On the practical side, it is difficult to use because of the phase-space cuts. On the conceptual side, the cuts introduce additional scales into the soft and jet functions which frustrate factorization and resummation. The most serious flaw, however, is that the jet and soft functions are not individually infrared safe: they each have infrared divergences which cancel only when combined, as we saw with the explicit example above. These divergences come from an incomplete cancellation between the real-emission graphs, which have phase-space restrictions, and the virtual graphs, which do not. We could attempt to put phase-space cuts on the virtual graphs as well. However, it is more logical to try to remove the phase-space cuts from the real-emission contributions to the jet and soft functions, since this would simplify their calculation and removes the spurious scales.

First we remove $\Lambda$. This is quite simple. Only the soft function depends on $\Lambda$. By our definition, $S^{\scL\scRb}$ in \Eq{factcut} only integrates over the soft-but-not-collinear region of phase space. The phase-space region outside of the cones but with energy $\Lambda < E < \infty$ does not contribute at all at leading power in $\tau$. So we can simply define a new soft function by including also this region:
\be
S^{\scRb} \,\LPeq\, S^{\scL \scRb}
\label{InfiniteLambda}
\ee
where $S^{\scRb} = S^{(\scL = \scinf)\scRb}$ has no cutoff on energy in the soft function. This equivalence can be verified at order $\alpha_s$ in Eq.~\eqref{softresExpResult}, where the entire $\Lambda$ dependence is subleading power in $\tau$, as observed above. Explicitly,
\begin{multline}
S^{ \scRb } (\tau)  =  \delta(\tau) + C_F \frac{ \alpha_s }{ \pi}  \left( \frac{\mu^2}{ Q^2} \right)^\e \Bigg\{   \delta(\tau) \left[ -\frac{1}{\e^2} - \frac{ 7 \pi^2}{12}  \right]
\\
- 2 \delta(\tau) \left[ - \frac{1}{\e} \ln \omega - \ln \omega \ln R+  \ln^2 \omega  +\cO(R) \right]  -  \bigg[\frac{2 }{\tau}  \ln R  \bigg]_ +  \Bigg\} 
\label{softresExpResultR}
\end{multline} 
Note that $S^\scRb$ is identical to the coefficient of $\theta(\Lambda-\frac{\tau}{R})$ in Eq.~\eqref{softresExpResult}. One might have imagined that taking $\Lambda \to \infty$ would introduce new UV poles. However, radiation in $\scRb$, say in the right hemisphere, at a 
given $\tau$ must have $k^+ = Q\tau$ and $k^- < \frac{1}{R} k^+ = Q\frac{\tau}{R}$, so the energy $E=\frac{1}{2}(k^+ + k^-)<\frac{Q\tau}{2}(1+\frac{1}{R})$ of all radiation
contributing to $S^\scRb$ at fixed $\tau$ is in fact bounded from above so there are no new UV divergences.  Thus, we now have
\be
 \frac{d\sigma}{d \tau}
\LPeq\, H \times S^\scRb \otimes  J^{\ccR_\ccO} \otimes J^{\ccR_\ccT} 
\ee
with no $\Lambda$ dependence on either side. Keep in mind that this equivalence is still valid only as $R\to0$: there are power corrections in $R$ on the right-hand side.

Removing the $R$ dependence is more subtle, since the $R$ dependence in both the soft and jet functions is relevant at leading power in $\tau$ and since the dependence on $R$ in
both functions is singular as $R\to0$. To remove it, we need a subtraction. 
To construct the subtraction, first recall that the general amplitude-level factorization proof in \Loop{} applies to any operator, including one composed of Wilson lines.
In particular, collinear factorization for a Wilson-line operator implies
\be
S \LPeq S^{\scRb}  \otimes J_\eik^{\ccR_\ccO} \otimes J_\eik^{\ccR_\ccT} 
\label{Seik}
\ee
where $S=S^{\overline{\scR = \infty}}$ has no angular or energy restriction and the  {\bf eikonal jet function} is defined as
\be
J^{\ccR_\ccO}_\eik(\tau) = \sum_{X_\ccO} \int d \Pi_{X_\ccO} 
\left| \frac{\bra{X_\ccO} Y^\dag_\ccO W_\ccO \ket{0}}{\bra{0} Y_\ccO^\dg W_\ccO \ket{0}} \right|^2
\delta (\tau-p_{X_\ccO}^+)
\label{jeteikcutdef}
\ee
The eikonal jet function differs from the jet function in Eq.~\eqref{jetcutdef} in that $Y_\ccO^\dag$ replaces the field $\bar{\psi}$. 
Note that the measurement function in the eikonal jet
function is the power-expanded version, $\delta(p-p^+)$, rather than $\Omega_\tau(p)$. This is consistent with Eq.~\eqref{Seik}
since the phase space in the eikonal jet function is restricted to be in a cone.

Explicitly, to order $\alpha_s$, we find
\begin{multline}
 J_\text{eik}^{\ccR_\ccj}(\tau) = \delta (k) + \frac{\alpha_{s} C_{F}}{ 2 \pi } \left(\frac{ \mu^{2} }{ Q^{2} }\right)^{\e}  \Bigg\{  \delta(\tau) \left[  \frac{2\pi^2}{3} -\frac{2}{\e} \ln{\omega} -2 \ln{\omega} \ln R +2\ln^2{\omega} +\cO(R)  \right]  \\ +  \left( \frac{2}{\e} + 2 \ln R \right) \left[\frac{1}{\tau} \right]_{+}  - 4 \left[ \frac{\ln{\tau}}{\tau}  \right]_{+} \Bigg\}
 \label{jeteikcutExpResult}
\end{multline}
Comparing \Eq{jeteikcutExpResult} to \Eq{jetcutExpResult}, we see that the $\omega$ and $R$ dependence in $J_\text{eik}^{\ccR_\ccj}$ is the same as that in $J^{\ccR_\ccj}$. This is expected, since the only IR-sensitive difference between the two is in the collinear-but-not-soft region of $\bra{X_\ccO} Y^\dag_\ccO W_\ccO \ket{0}$ and $\bra{X_\ccO} \bar\psi W_\ccO \ket{0}$.
In this region,  there is a complete cancellation of real and virtual graph for both functions, hence both are IR-finite.
Note also that there are no $\frac{1}{\e^2}$ poles in the eikonal jet function. These double UV poles in the regular jet function
come from virtual graphs. In the eikonal jet function, the virtual graphs in the numerator and denominator of Eq.~\eqref{jeteikcutdef}
are identical and hence cancel in the ratio to order $\alpha_s$. The lack of $\frac{1}{\e^2}$ poles also implies
that there are no Sudakov double logs in the eikonal jet function.

Now, if we convolve both sides of Eq.~\eqref{factcut} with the eikonal jet functions and use \Eq{InfiniteLambda} and \Eq{Seik}, we get
\be
 \frac{d\sigma}{d \tau}\conv J_\eik^{\ccR_\ccO} \conv J_\eik^{\ccR_\ccT} \LPeq H \times  S 
 \conv J^{\ccR_\ccO}
 \conv J^{\ccR_\ccT}
\label{Swithe}
\ee
At this point, no object in this leading-power equivalence depends on $\Lambda$ and the $R$ dependence on both sides is only in the jet functions and eikonal jet functions. We still must have $R$ small though, since there are power corrections in $R$ on both sides.

Finally, we want to remove the $R$-dependence completely. Let us call a jet function
with no restriction on $R$ an {\bf inclusive jet function} and denote it by $J^\ccj$. 
Removing the $R$ introduces additional unphysical singularities collinear to the Wilson-line direction $t_\ccj$
which are not regulated with the off-shellness regulator. We must introduced another regulator for these singularities,
so we use the $\Delta$-regulator \cite{Chiu:2009yx}, $\delta_\ccj$, as shown in Eq.~\eqref{deltareg}.
To order $\alpha_s$ we find for the inclusive jet function
\begin{multline}
J^\ccj(\tau ) = \delta(\tau)+ \frac{\alpha_{s} C_{F}}{ 2 \pi } \left(\frac{ \mu^{2} }{ Q^{2} }\right)^{\e}  \bigg\{  \delta(\tau)  \left( \frac{2}{\e^2} + \frac{3}{2 \e} + \frac{7}{2} -\frac{\pi^2}{6}   \right) 
\\
 +\delta(\tau) \left( -\frac{2}{\e} \ln{\omega} +2 \ln{\omega} \ln \delta_\ccj + \ln^2{\omega}   \right)
 - \left( 2 \ln \delta_\ccj + \frac{3}{2}  \right) \left[\frac{1}{\tau} \right]_{+} \bigg\} 
 \label{JjExpResult}
\end{multline}
Similarly, for the inclusive eikonal jet function we find
\begin{multline}
 J^\ccj_\text{eik} (\tau ) = \delta(\tau)+ \frac{\alpha_{s} C_{F}}{ 2 \pi } \left(\frac{\mu^{2} }{ Q^{2} }\right)^{\e}  \bigg\{ \delta(\tau) \left[ \frac{\pi^2}{3} -\frac{2}{\e} \ln{\omega} + 2\ln{\omega} \ln \delta_\ccj+ \ln^2{\omega}  \right]  
\\
+  \left( \frac{2}{\e} - 2 \ln \delta_\ccj \right) \left[\frac{1}{\tau} \right]_{+}  - 2 \left[ \frac{\ln{\tau}}{\tau}  \right]_{+} \bigg\}
 \label{JjeikExpResult}
\end{multline}
Note that the $\delta_\ccj$ dependence associated with the Wilson line direction is identical in the two inclusive jet functions.

Next, note that since  $\overline{\ccR_\ccj}$ does not contain the jet direction  the only leading-power contributions to the
jet function from this region are soft.\footnote{One might be concerned about collinear singularities associated with the Wilson-line direction $t_\ccj^\mu$. However, since  the measurement function forces
$p\cdot n = \tau$, at small $\tau$ radiation cannot be collinear to both $t_\ccj^\mu$ and the jet direction $n^\mu$. Thus, radiation collinear to $t_\ccj^\mu$ cannot contribute at leading power.}
Thus, we can apply the general amplitude-level factorization theorem 
to the operator $\bar{\psi} W_\ccj$ to get
\be
J^\ccj \,\LPeq\, J^{\ccR_\ccj} \otimes J_\text{eik}^{\overline{\ccR_\ccj}}
\label{jdecomp}
\ee
This equation can be verified at 1-loop by comparing \Eq{JjExpResult} with the combination of Eqs.~\eqref{jetcutExpResult}, \eqref{jeteikcutExpResult} and \eqref{JjeikExpResult}.
Similarly,
\be
J^\ccj_\eik \,\LPeq\, J^{\ccR_\ccj}_\eik \otimes J_\text{eik}^{\overline{\ccR_\ccj}}
\ee
Therefore, convolving both sides of \Eq{Swithe} with $J_\text{eik}^{\overline{\ccR_\ccO}}$ and  $J_\text{eik}^{\overline{\ccR_\ccT}}$ gives
\be
 \frac{d\sigma}{d \tau} \conv J^\ccO_{\eik} \conv J^\ccT_\eik \LPeq H \times  
S
 \conv J^\ccO
 \conv J^\ccT
\label{Swithe2}
\ee
In this final form, all the dependence on $\Lambda$ or $R$ has been explicitly removed.

Finally, we want to isolate $\frac{d \sigma}{d\tau}$ from Eq.~\eqref{Swithe2}. To do this, we use that convolutions map to products in Laplace space. Taking the Laplace transform, Eq.~\eqref{Swithe2} translates to
\be
 \int d \tau\frac{d\sigma}{d \tau} e^{-\nu \tau}
 \,\LPeq\, H \, \frac{\widetilde S(\nu) \widetilde J^\ccO (\nu) \widetilde J^\ccT (\nu)}
{\widetilde J^\ccO_\eik (\nu) \widetilde J^\ccT_\eik (\nu)}
\ee
This form is in agreement with previously known expressions in the literature \cite{Berger:2003iw,Lee:2006nr}.

\subsection{Jet broadening}

The above discussion shows how the phase-space cutoffs separating collinear and soft radiation as well as the UV phase-space cutoff can be removed in the factorization formula for a particular observable (thrust). The derivation easily generalizes to many other observables. The key general property that was used is that the vanishing limit of the observable forces the phase space into the $N$-jet configuration at leading power. This allows the factorization theorem in \Eq{QEDmain} to be used to factorize the matrix-element squared in the full distribution. It also ensures that the dependence on the phase-space cutoffs is power suppressed, once the eikonal jet functions are included. For observables whose measurement function is not linear in each sector, the integrals will not be a simple convolution.

For a marginally different example, consider jet broadening~\cite{Rakow:1981qn,Ellis:1986ig,Catani:1992jc,Chiu:2011qc,Becher:2011pf}. (Total) jet broadening acting on a state $\ket{X}$ with particles of momenta $p_j^\mu$ has the eigenvalue
\be
b(X) = \frac{1}{2Q} \sum_j \Omega_b(p_j),\qquad \Omega_b (p) = |\vec{p}_{\perp}|
\ee
where  $\vec{p}_{\perp}$ are the components of the 3-momenta of the particles transverse to the thrust axis. 

In the SCET literature, jet broadening is considered a SCET${}_\mathrm{II}$ observable because soft emissions which are hard enough to recoil against collinear emissions
contribute to jet broadening at leading power, while they are subleading power for thrust. More explicitly, for thrust only the small component of momentum $p^+ = n \cdot p$ contributes (for particles going in the $n$ hemisphere). Thus collinear momenta, with $(p^-, p^+, p^\perp) \sim Q(1,\lambda,\lambda^2)$ and soft momenta with  $p \sim Q \lambda^2$ contribute at the same order. Soft momenta scaling like $p \sim Q \lambda$ give a power-suppressed contribution to thrust. For jet broadening, $p_\perp$ is measured. So collinear momenta
contribute $p_\perp\sim Q\lambda$ and therefore soft momenta scaling like $p\sim Q\lambda$ are relevant at leading power making jet broadening a SCET${}_\mathrm{II}$ observable.

From the point of view of the factorization as set up in \Tree{} and \Loop{}, the soft scaling is unrelated to the collinear scaling. That is, the amplitude-level factorization formula, Eq.~\eqref{QEDmain} holds
for any relationship between the soft-scaling parameter, $\lambda_\scs$, and the collinear-scaling parameter, $\lambda_\ccc$. SCET${}_\mathrm{I}$ corresponds to $\lambda_\scs = \lambda_\ccc^2$ and SCET${}_\mathrm{II}$ to $\lambda = \lambda_\ccc$. The relevant implication of the soft and collinear momenta having $p_\perp$ components of the same order 
is that configurations where soft particles recoil against collinear particles must be accounted for in the factorization theorem. The result is that the factorization formula has the form
\be
\frac{ d \sigma}{ d b}
 = H \int d b_\scs d b_{\ccO} d b_{\ccT} d^2  p_\ccO^\perp d^2 \vec p_\ccT^\perp J^\ccO (b_\ccO,\vec p_\ccO^\perp) J^\ccT (b_\ccT,\vec p_\ccT^\perp)  S(b_\scs,-\vec p_\ccO^\perp, - \vec p_\ccT^\perp)
\delta(b-b_\scs - b_\ccO - b_\ccT)
\label{fb}
\ee
We can write this heuristically as
\be
 \frac{ d \sigma}{ d b} 
\LPeq H \times J^\ccO \otimes J^\ccT \otimes S\label{fbfact}
\ee
with the understanding that $\otimes$ for jet broadening refers to the double convolution in Eq.~\eqref{fb}.

With phase-space restrictions, these jet functions are given by
 \be
J^{\ccR_\ccO} (b,\vec p^\perp ) = \sum_{X_\ccO} \int d \Pi_{X_\ccO} 
\left| \frac{\bra{X_\ccO} \bar\psi W_\ccO \ket{0}}{\bra{0} Y_\ccO^\dg W_\ccO \ket{0}} \right|^2
\delta \Big(b-b(X_\ccO) \Big) \delta(Q- p_{X_\ccO}^+) \delta(\vec p_\perp - \vec p_{X_\ccO}^\perp)
\label{broadjetcutdef}
\ee
and the soft function by
 \begin{multline}
S^{\scL \scRb} (b, \vec p^\perp_\ccO, \vec p^\perp_\ccT ) = \sum_{X_\scs} \int d \Pi_{X} 
\left|  \bra{X_\scs} Y_\ccO^\dg Y_\ccT \ket{0}  \right|^2
\delta \Big(b-b(X_\scs)\Big) \left[
\delta(\vec p^\perp_\ccO - \vec p_{X_\scs}^\ccO) + \delta(\vec p^\perp_\ccT - \vec p_{X_\scs}^\ccT)
\right]
\label{broadsoftcutdef}
\end{multline}
where $\vec p_{X_\scs}^\ccO$ is the net $\perp$ momenta in the left hemisphere and $\vec p_{X_\scs}^\ccT$ is the net $\perp$ momenta in the right hemisphere.
As with thrust, these phase-space restricted functions will have overlapping UV-IR divergences and unwieldy dependence on the cutoffs $R$ and $\Lambda$. However, as with thrust, we 
can convolve both sides of Eq.~\eqref{fbfact} with eikonal jet functions to get a factorization formula with only objects with no phase space cutoffs. The result in
Laplace space is
\be
 \int d b\frac{d\sigma}{d b} e^{-\nu b}
 \cong
  H \, \int \rd^{2} x_{L}^{\perp} \rd^{2} x_{R}^{\perp}\, \frac{ \wtd{J} (x_{L}^{\perp}, \nu )  \,  \wtd{J} (x_{R}^{\perp}, \nu ) \,  \wtd{S} (x_{L}^{\perp} , x_{R}^{\perp}, \nu )   }{ \wtd{J}_\text{eik} (x_{L}^{\perp} ,\nu ) \,  \wtd{J}_\text{eik} (x_{R}^{\perp} , \nu)   }
\ee
Here, $\nu$ is the Laplace-conjugate variable to $b$ and $x_\ccO^\perp$ and $x_\ccT^\perp$ are the Laplace conjugate variables to $p_\ccO^\perp$ and $p_\ccT^\perp$ respectively.

\subsection{Comparison to other approaches}
\label{sec:compare}

We have seen how phase-space cutoffs can be removed for certain inclusive observables if the double counting in the soft-collinear region is
removed with an eikonal jet function.
 In this section,
we would like to emphasize some conceptual differences in our derivation and previous ones and contrast with the literature.

First of all, it is easy to compare our results to those in traditional QCD, where the eikonal jet function first appeared. The final factorization formulas
are identical. One difference is that in the early literature the eikonal jet functions were subtracted from the soft function rather than the jet function. From the point of
view of the final formula, there is no difference. However, conceptually our analysis makes it clear that the eikonal jet function should be subtracted from the jet function rather than
the soft function. Indeed, the soft function is by itself infrared finite while the naive inclusive jet function (without the subtraction) is not. As shown explicitly in Eq.~\eqref{JjExpResult}, the infrared
divergences do not cancel between the real and virtual graphs for the jet function. With the subtraction the jet function is a well-defined and infrared safe object. The fact that
the subtraction is more naturally applied in the collinear sector was also appreciated in~\cite{Lee:2006nr}.

The comparison to SCET is perhaps more illuminating than the comparison to traditional QCD. 
In the early days of SCET, calculations were mostly done in dimensional regularization (DR) and the overlapping of soft and collinear phase-space regions 
were not much discussed.
In retrospect, it is easy to see why the correct answers result in DR without a subtraction: the eikonal jet functions, as in Eq.~\eqref{jeteikcutdef}, give scaleless integrals in DR
and thus formally vanish and can be ignored.

It is natural to be somewhat uncomfortable with setting scaleless but IR and UV divergent integrals to zero in DR. The mathematical justification notwithstanding, it is dangerous from a practical point of view if one hopes to extract an anomalous dimension from the poles in the jet and soft functions at $d=4$. The only way it will work is if the object one computes is infrared finite. For infrared finite objects, all the poles are by definition  UV. So setting $\frac{1}{\euv}-\frac{1}{\eir} =0$ has no effect. As we have shown, the subtracted inclusive jet function is IR finite,
so practically, one can ignore the subtraction in DR. Morally, though, to do this one must be able to show that the jet function is IR finite. Without the subtraction it is not. 
In this respect, the success of SCET in pure DR was somewhat accidental.

The missing subtractions were understood in the classic paper on zero-bin subtraction by Manohar and Stewart~\cite{Manohar:2006nz}. These authors showed the the proper derivation of the effective Lagrangian for SCET involves binning the momenta into collinear momenta in different directions and soft momenta. The zero bin in each collinear sector should be formally excluded. In~\cite{Manohar:2006nz} it was shown that this exclusion amounts to the subtraction diagram-by-diagram of the soft-limit of the collinear momenta. In~\cite{Lee:2006nr,Idilbi:2007ff,Idilbi:2007yi}, this subtraction procedure
was shown to be equivalent to the eikonal jet function subtraction of traditional QCD.

A somewhat different perspective comes from the method-of-regions multipole-expansion approach to
SCET~\cite{Beneke:1997zp,Beneke:2002ph,Beneke:2002ni}, as recently reviewed in~\cite{Becher:2014oda}. In this approach, the soft-collinear subtraction and the extension of the soft integrals to $\Lambda=\infty$ are not discussed or needed. The basis of the argument is that the non-analytic dependence on the observable in each region is independent of the possible phase-space cutoffs. Since all the physics
is in this non-analytic dependence, one can remove the cutoffs without consequence. For more details, see~\cite{Beneke:1997zp,Becher:2014oda}.
\newpage

\section{Conclusion}
\label{sec:conclusions}

In this paper we have presented two new results. First, we have given a recursive formula for constructing  amplitudes which agree with full QCD at leading power to all-loop order in any soft and collinear limit. The
subtracted amplitudes we describe are matrix elements of fields and Wilson lines. Unlike with the amplitude-level factorization formula in \Loop{}, one does not have to specify whether the particles are soft or collinear ahead of time: the subtracted matrix elements will be correct in any limit. Although the amplitudes appear simpler than in full QCD (for example, the only interference effects from different directions involve gluons emitted off Wilson lines), it remains to be seen whether they can be integrated simply to provide a productive subtraction scheme. In our
derivation of this formula, extensive use was made of the proof of factorization in \Loop{}.

Second, we showed how phase-space cutoffs can be removed when integrating a factorized amplitude squared against the measurement function for certain inclusive observables. Removing the cutoffs does two things: it overcounts the soft-collinear region and adds UV divergences to the phase-space integrals. These two effects can be compensated for by integrating the full QCD distribution against an eikonal jet function. This convolution can be easily disentangled, at least for thrust, jet broadening and angularities. 
This extends the results of amplitude-level factorization from \Tree{} and \Loop{} to the level of observables.

In our presentation, we have included explicit 1-loop expressions for soft and jet functions with cutoffs and for the eikonal jet function in a regularization scheme which separates UV from IR. 
These expressions confirm generally the qualitative analyses that we have presented of the UV and IR structure of the integrals.

 Although our final factorization formulas are not new, we believe our derivation is systematic and rigorous. 
We hope that the step-by-step procedure we have presented will be useful in future studies of factorization, where subtleties abound.

\label{sec:Conclusions}

\section{Acknowledgements}
\label{sec:Acknowledgements}
The authors would like to thank Thomas Becher, Christopher Lee and Hua-Xing Zhu for useful conversations.
The authors are supported in part by grant DE-SC003916 from the Department of Energy.

\section{Appendix}

In this appendix, we report the explicit computations of many of the cut-off dependent objects that appear in Section~\ref{sec:CrossSectionThrust}. All expressions are computed at one-loop order in QCD for the observable thrust.

\subsection{Regularization schemes}
We will use dimensional regularization to control the UV divergences: we analytically continue to $d=4-2\e$ dimensions, with $\e>0$. 

To regulate the IR divergences, we take the outgoing external fermion lines 
to have an offshellness of $Q^2 \omega$. Consequently, the propagator in $Y_\ccj$ will look like 
\be 
 \frac{n_\ccj^\mu}{n_\ccj \cdot k } \rightarrow \frac{p_\ccj^\mu}{ p_\ccj \cdot k+ \frac{Q^2 \omega}{2} }  
\label{omegareg}
\ee
It is helpful when using an offshellness also to slightly modify the measurement function for thrust, so that virtual graphs
still contribute only at $\tau=0$. We can do this by replacing $\delta(\tau - \frac{1}{Q} p^-) \to \delta(\tau - \frac{1}{Q} p^- + \omega)$.

In scaleless integrals involving Wilson lines, the offshellness may not completely control all the IR divergences. 
Thus, in addition we use the $\Delta$ regulator \cite{Chiu:2009yx} for the 
 collinear Wilson line, so that the propagator in the $W_\ccj$  will be shifted by $  \frac{ \Delta }{  t_{\ccj} \cdot p_{\ccj}}  $ .
More precisely, we define the dimensionless parameter $\delta_\ccj \equiv \Delta/(t_\ccj \cdot p_\ccj)$ 
and shift the eikonal propagators as
\be  \label{deltareg}
\frac{1}{t_{\ccj} \cdot k} \rightarrow \frac{1}{t_{\ccj} \cdot k + \frac{ \Delta }{  t_{\ccj} \cdot p_{\ccj}}  }    \equiv   \frac{1}{t_{\ccj} \cdot k + \delta_{\ccj}  ( t_\ccj \cdot p_\ccj  )}
\ee
With these two IR regulators,  all the $1/\e$ poles in the following expressions correspond to UV divergences only.

For simplicity, in the following sections we choose the direction of the collinear Wilson line, $W_{\ccO}$ to be $t_\ccO^\mu=n_\ccT^\mu = (1,-\vec n)$ and that of $W_\ccT$ to be $t_\ccT^\mu= n_\ccO^\mu = (1,\vec n)$, where $\vec n$ is the thrust axis.  Then, in light-cone coordinates
\be
k^\mu = \frac{1}{2} k^-  n_\ccO^\mu + \frac{1}{2} k^+ n_\ccT^\mu + k_\perp^\mu
	= (k^-, k^+, k_\perp)
\ee
While it is not impossible to do the integrals for generic choices of $t_\ccO^\mu$ and $t_\ccT^\mu$, the phase space integrals inside the
cone become significantly more complicated.

\subsection{Jet Functions \label{sec:jets}}
We will denote a jet function restricted to a cone of size $R$ by $J^\ccR$. In our notation, $R$ is a rapidity-type cone angle:
a particle is in the $n$-cone if $k^+ < R k^-$. $R$ is related to
the cone opening angle $\theta$ by $R = \tan^2\frac{\theta}{2}$. 

At order $\alpha_s$, the virtual contributions to the jet functions are of course independent of phase space restrictions and therefore the same for the restricted
or unrestricted jet functions. The purely virtual contributions come from evaluating
\begin{align}
 \label{VirtJRes}
 \Big[J(\tau)\Big]_\text{virt.}  &=  \int \frac{\rd p^+}{ 2 \pi} 
\left| \frac{\bra{p} \bar\psi W_\ccj \ket{0}}{\bra{0} Y_\ccj^\dg W_\ccj \ket{0}} \right|^2 2 \pi \delta \Big ( Q p^+ - Q^2 \omega\Big)  \theta(p^+)
\delta\big(\tau  - \frac{p^+}{Q}  + \omega)\\
&= \delta(\tau)\frac{1}{Q} \left| \frac{\bra{p} \bar\psi W_\ccj \ket{0}}{\bra{0} Y_\ccj^\dg W_\ccj \ket{0}} \right|^2
= \delta(\tau) + \cO(\alpha_s)
\end{align}
where the sum over spins is implicit. 

There are 3 virtual graphs contributing to $\bra{p_\ccj} \bar\psi W_\ccj \ket{0}$ at 1-loop. The self-energy graph
on the Wilson line exactly vanishes; the vertex correction and quark self-energy graph sum to
\begin{multline}
\frac{\bra{p} \bar \psi W_\ccj \ket{0}_\text{virt.}}{\bra{p} \bar \psi W_\ccj \ket{0}_\text{tree}} =  g_s^2 \mu^{2 \e} C_F  \int \frac{ \rd^{d} k }{ (2 \pi)^d }
 \left( \frac{2 (p^- - k^-) }{k^+ + p^- \delta_\ccj } + \frac{d-2}{2}   \frac{ p^+  - k^+  }{ p^+ }   \right)  \frac{1}{ k^2 } \frac{1}{(p^- - k^-) (p^+ - k^+)- k_\perp^2}\\
= 
-
\frac{ \alpha_s C_F}{4 \pi} \left( \frac{4 \pi \mu^2}{ Q^2} \right)^\e  (-\omega)^{-\e} \Gamma(\e) \,  \frac{\Gamma(1-\e) \Gamma(2-\e) }{ \Gamma(2-2\e)}\\
\hspace{4cm}\times \left[    -(3+\e) + 4\Big(1+\frac{1}{\delta_{\ccj}} \Big)\,{}_{2}F_{1} (1, 1-\e, 2-2\e, -\frac{1}{\delta_{\ccj}}) \right]    \\
=
 -
 \frac{ \alpha_s C_F}{4 \pi } \left( \frac{\wtd \mu^2}{ Q^2} \right)^\e \left[ \frac{3}{2 \e} - \frac{3}{2} \ln \omega 
 +   \frac{2}{\e} \ln \delta_\ccj - 2\ln \omega \ln \delta_\ccj -  \ln^2 \delta_\ccj+ \frac{7}{2} - \frac{2 \pi^2}{3}\right]
\label{Jvtone}
\end{multline}
In the last step, we expanded in $\delta_\ccj$ and $\omega$ and dropped the $\cO(\delta_\ccj) $ and $\cO(\omega) $ terms.

To one-loop, the denominator factor in the jet function evaluates to
\begin{multline}
\IRZ_\ccj =  \bra{0} Y_\ccj^\dg W_\ccj\ket{0} =   1+ g_s^2 \mu^{2 \e} C_F  \int \frac{ \rd^{d} k }{ (2 \pi)^d }  \, 2 \left( \frac{1}{- k^+ + Q \omega } \frac{1}{k^- + Q \delta_\ccj }   \right) \frac{1}{k^2}
\\
=  -
 \frac{ \alpha_s C_F}{4 \pi} \left( \frac{4 \pi \mu^2}{ Q^2} \right)^\e ( -\omega)^{-\e}  \Gamma(\e) \Big[ 2 \delta_\ccj ^{-\e} \Gamma(\e) \Gamma(1-\e) \Big]   \\
 = - 
 \frac{ \alpha_s C_F}{4 \pi }  \left( \frac{\wtd \mu^2}{ Q^2} \right)^\e \left[ -\frac{2}{\e^2} + \frac{2}{\e} \ln \omega - \ln^2 \omega 
  +  \frac{2}{\e} \ln \delta_\ccj - 2\ln \omega \ln \delta_\ccj - \ln^2 \delta_\ccj - \frac{ \pi^2}{ 2}   \right]
\label{Zone}
\end{multline}
We note here that  the $\delta_\ccj$ dependence is identical in Eqs.~\eqref{Jvtone} and \eqref{Zone}. Thus
virtual IR divergences introduced by the $W_\ccj$, which are regulated with the $\Delta$-regulator, cancel in $[J(\tau)]_\text{virt.}$
to one loop. Explicitly, 
\begin{align}
 \Big[J(\tau)\Big]_\text{virt.}  &= \delta(\tau)-\delta(\tau) \frac{ \alpha_s C_F}{2 \pi }  \left( \frac{\wtd \mu^2}{ Q^2} \right)^\e
\left[
\frac{2}{\e^2} +
\frac{3}{2 \e}- \frac{2}{\e} \ln \omega - \frac{3}{2} \ln \omega + \ln^2 \omega + \frac{7}{2} - \frac{\pi^2}{6} 
\right]
\end{align}

Now let's look at the real-emission diagrams. With emissions restricted to a cone of size $R$,
the real-emission contributions to the jet function at order $\alpha_s$ are
\begin{multline}
\Big[ J^{\ccR_\ccj} (\tau) \Big]_{\text{real}} =
  2 g_s^2 \mu^{2 \e} C_F  \int  \frac{\rd p^+ }{ 2 \pi} \int  \frac{\rd p^- }{ 2 \pi} (2\pi)\delta(p^- - Q)  \\
\times 
\int \frac{ \rd^{d} k }{ (2 \pi)^d }   \frac{1}{p^+}\left( \frac{2 (p^- - k^-) }{k^+ + p^- \delta_\ccO } + \frac{d-2}{2}   \frac{ p^+  - k^+  }{ p^+  }   \right)
  2 \pi \delta(k^+ k^- - k^2_\perp) \theta (k^+) \theta (k^-) \\ 
\times  2\pi \delta\Big( (p^+  - k^+) (p^- - k^-) -  k_\perp^2 - Q^2 \omega  \Big) \theta(p^+ - k^+ )\theta(p^- - k^- )  \delta(\tau - p^+ /Q  + \omega)\\
\times \theta \big( \frac{k^-}{k ^+}  -  R \big) \theta \big(\frac{ p^- - k^-}{ p^+ - k^+ }- R \big)
\hspace{5cm}
\label{Jwithphase}
\end{multline}
It is helpful to write the result as
\be
\Big[ J^{\ccR_\ccj} (\tau) \Big]_{\text{real}}
=  \frac{ \alpha_s C_F}{2 \pi} \frac{1}{\Gamma(1-\e)} \left( \frac{4 \pi \mu^2}{ Q^2} \right)^\e 
 \frac{1}{\tau+\omega}\frac{1}{\tau^\e} \, \cI^{\ccR_\ccj} (\tau)
\label{helpful}
\ee 
and we find
\begin{align}
\cI^{\ccR_\ccj}(\tau) &= \int_{ \tau / R}^1 \rd x \, \left[ \frac{2(1-x)}{x+ \delta_\ccj}   +(1-\e) x  \right] \left[ \frac{1}{x(1-x)} \right]^\e   
\notag\\&=  
\cI^\ccj - \frac{2  (1+\delta_\ccj) }{ (1- \e) \delta_\ccj} \left(  \frac{\tau}{R}  \right)^{1-\e} F_1\left(1-
\e, \e,1; 2-\e;  \frac{\tau}{R}  ,  -\frac{ \tau }{R \, \delta_\ccj} \right)  
\notag\\& \hspace{3cm} + \frac{1}{2} \left[ \frac{\tau}{R} \left(1- \frac{\tau}{R} \right) \right]^{1-\e} + \frac{1}{2} (3+\e) B \left( \frac{\tau}{R} ; 1-\e, 1-\e \right) 
\notag\\&= 
\left( - \frac{3}{2} - 2 \ln \Big( \frac{\tau}{R} +  \delta_\ccj \Big)  + \cO(\tau)  \right) + \cO(\e) 
\end{align}
where $F_1(\alpha, \beta, \beta' ; \gamma; x, y)$ is the Appell hypergeometric function and $B(z; a, b)$ is the Incomplete Beta Function.
These real emission graphs are of course UV finite, so we can simply set $\e =0$. 

To expand the result for small $\omega$, we can use that in the limit $\omega\to 0$
\be
 \frac{1}{\tau+\omega} \, \cI (\tau)=\delta(\tau)\left[\int_0^1  \frac{d \tau'}{\tau' + \omega }   \cI (\tau')\right]
+\left[\frac{\cI(\tau)}{\tau}\right]_+
\label{Itodist}
\ee
so that
\begin{multline} \label{JRj}
\Big[ J^{\ccR_\ccj} (\tau) \Big]_{\text{real}}
 =  \frac{ \alpha_s C_F}{2 \pi } 
 \Bigg\{ \delta(\tau) \left[  \frac{3}{2} \ln \omega - 2  \ln \omega \ln R + \ln^2 \omega + \cO(R) \Big) \right] +  \left[ \frac{1}{\tau} \left( - 2 \ln \frac{\tau}{R} - \frac{3}{2} \right)\right]_+ \Bigg\}
\end{multline}
\Eq{JRj} shows that the jet function in cone does not have $\log \delta_\ccj$ singularity.
Note, however, that this jet function does have double logs of $\omega$, coming from soft-collinear region; these
will cancel against the virtual soft-collinear singularity in Eq.~\eqref{Zone}.
The $R$ dependent terms $\log \omega \log R$ come from soft emissions at the cone edge. These do not cancel against any other contribution
to $J^\ccR$, but will cancel against either $S^\scRb$ of the eikonal jet function $J_\text{eik}^\ccR$.

Next, consider the inclusive jet function, $J^\ccj$. This is defined identically to $J^\ccR$, but without the phase space restriction.
The virtual contributions are the same. The real emission contributions at order $\alpha_s$ are the same without the phase space restriction
on the last line of Eq.~\eqref{Jwithphase}. Using the same notation as in Eq.~\eqref{helpful}, we find
\begin{align}
\cI^\ccj &= \int_0^1 \rd x \, \left[ \frac{2(1-x)}{x+ \delta_\ccj}   +(1-\e) x  \right] \left[ \frac{1}{x(1-x)} \right]^\e  \notag\\
&=  \frac{4^{-1+\e} \sqrt{\pi} \Gamma(1-\e)}{ \delta_{\ccj} \, \Gamma(\frac{3}{2}-\e)} \left[    -(3+\e)\delta_{\ccj} + 4(1+\delta_{\ccj})\,  {}_{2}F_{1} (1, 1-\e, 2-2\e, -\frac{1}{\delta_{\ccj}}) \right]  
\\
&= -\frac{3}{2} + 2(1+\delta_\ccj)\ln(1+\frac{1}{\delta_\ccj})
\end{align}
We have set $\e=0$ on the last line since the real emission contribution to the inclusive jet function, like with $J^\ccR$, is UV finite.
Using Eq.~\eqref{Itodist} we then find
\be
\Big[ J^{\ccj} (\tau) \Big]_{\text{real}}
 =  \frac{ \alpha_s C_F}{2 \pi }   \Bigg\{ \delta(\tau) \left(  \frac{3}{2} \ln \omega + 2  \ln \omega \ln \delta_\ccj \right) +  \left[ \frac{1}{\tau} \left(- 2 \ln \delta_\ccj - \frac{3}{2} \right)\right]_+ \Bigg\}
\ee
The phase space integral in the inclusive jet function contains single log of $\omega$, indicating pure collinear singularity. It also contains a double IR singularity, $\log \omega \log \delta_\ccj $, coming from the soft-collinear region. 

Next, we consider the eikonal jet function, with the field $\bar{\psi}$ replaced by a soft Wilson line. The virtual contributions are given by
matrix elements similar to Eq.~\eqref{VirtJRes}:
\begin{align}
 \Big[J_\text{eik}(\tau)\Big]_\text{virt.}  &=  \int \frac{\rd p^+}{ 2 \pi} 
\left| \frac{\bra{0} Y_\ccO W_\ccO \ket{0}}{\bra{0} Y_\ccO^\dg W_\ccO \ket{0}} \right|^2 2 \pi \delta \Big (  p^+ - Q \omega\Big)  \theta(p^+)
\delta\big(\tau  -\frac{ p_\ccO^+ }{Q}  + \omega)\\
&= \delta(\tau)
\end{align}
Since there is an exact cancellation between numerator and denominator, the purely virtual contribution is $\delta(\tau)$ to all orders.

Using the same notation as above, for the phase-space restricted eikonal jet function, we find
\be
\cI_\text{eik}^{\ccR_\ccj}(\tau) = \int_{\tau /R }^\infty \rd x \, \frac{2}{x+ \delta_\ccj}  \left[ \frac{1}{x} \right]^\e  
= \delta_\ccj^{-\e} B\left(  \frac{R\, \delta_\ccj}{ \tau }; \e, 0 \right)   = \frac{2}{\e} - 2 \ln \left(\frac{\tau}{R}+ \delta_\ccj \right) + \cO(\e)
\ee
Note that now there is a UV divergence from the integral up to infinite energy in the $k^-$ direction.
This turns into
\begin{multline} \label{JeikRj}
\Big[ J_\text{eik}^{\ccR_\ccj}  (\tau) \Big]_\text{real}  
 =  \frac{ \alpha_s C_F}{2 \pi } \left( \frac{ \wtd \mu^2}{ Q^2} \right)^\e \\
\times 
\Bigg\{  \delta(\tau) \left[ \frac{2 \pi^2}{3}   - \frac{2}{\e} \ln{\omega} - 2\ln{\omega} \ln R + 2 \ln^2{\omega}  + \cO(R) \right] +   \left[\frac{1}{\tau} \left( \frac{2}{\e} + 2 \ln R - 4\ln \tau \right) \right]_{+} \Bigg\}
\end{multline}

For the unrestricted eikonal jet function
\be
\cI_\text{eik}^\ccj = \int_0^\infty \rd x \, \frac{2}{x + \delta_\ccj}  \left[ \frac{1}{x} \right]^\e  = \delta_\ccj ^{-\e} \,2\Gamma(\e) \Gamma(1-\e) 
\ee
and
\begin{multline}
\Big[J_\text{eik}^\ccj (\tau) \Big]_\text{real} =  
  \frac{ \alpha_s C_F}{2 \pi } \left( \frac{ \wtd \mu^2}{ Q^2} \right)^\e  \\
\times 
\Bigg\{ \delta(\tau) \left( \frac{\pi^2}{3}   - \frac{2}{\e} \ln \omega  +2  \ln \omega \ln \delta_\ccj +  \ln^2 \omega \right) +  \left[ \frac{1}{\tau} \left( \frac{2}{\e} - 2 \ln \delta_\ccj -2 \ln \tau  \right)\right]_+ \Bigg\}
\end{multline}

Finally, the jet functions with real and virtual contributions combined are given in Eqs.~\eqref{jetcutExpResult}, \eqref{jeteikcutExpResult}, \eqref{JjExpResult} 
and \eqref{JjeikExpResult} for $J^{\ccR_\ccj}$, 
$J_\text{eik}^{\ccR_\ccj}$, $J^\ccj$ and $J_\text{eik}^{\ccj}$ respectively. 
The structure of IR singularities in virtual and real contributions to the various objects are listed in Table \ref{IRsingular}.

\begin{table} 
\begin{center}
    \begin{tabular}{| c || c || c | c | c || c |c || c | c| }
    \hline
       &  {\ccol collinear} or {\softcol soft}  & \multicolumn{3}{ |c| }{ {\softcol soft}-{\ccol collinear} }  
       &\multicolumn{2}{ |c| }{ UV-IR } 
       &\multicolumn{2}{ |c| }{ UV } \\   \cline{2-9}
           & $\ln \omega $ & $\ln \omega \log R $ & $ \ln \omega \ln \delta_\ccj $ & $ \ln^2 \omega $ & $\frac{1}{\e } \ln \omega $ 
& $\frac{1}{\e} \ln \delta_\ccj$
& $\frac{1}{\e}$ & $\phantom{\dfrac{x^2}{2}} \frac{1}{\e^2}$
     \\  \hline \hline
$\phantom{\dfrac{x^2}{2}}\bra{p_\ccj} \bar \psi W_\ccj \ket{0}_\text{virt.}$
  & \checkmark  &  $-$ & \checkmark  & $-$  & $-$ & \checkmark  & \checkmark& $-$
     \\ \hline
$\phantom{\dfrac{x^2}{2}}\IRZ=\bra{0} Y_\ccj W_\ccj \ket{0}$
 & $-$   &  $-$ & $\checkmark$  & \checkmark   & \checkmark & \checkmark  & $-$ & \checkmark
     \\ \hline
            $ \phantom{\dfrac{x}{2}}\big[J \big]_\text{virt} $ & \checkmark  &  $-$ & $-$   &  \checkmark  & \checkmark &  $-$  &\checkmark&\checkmark
\\
\hline
\hline
       $ \phantom{\dfrac{x}{2}}\big[J^{\ccR_\ccj} \big]_\text{real} $  & \checkmark &  \checkmark   & $-$  & \checkmark   & $-$ & $-$ &$-$ &$-$ 

    \\[1mm] \hline 
   $ \phantom{\dfrac{x}{2}}\big[J^\ccj \big]_\text{real} $  & \checkmark &  $-$ & \checkmark  & $-$   & $-$ & $-$&$-$ &$-$ 
    \\[1mm] \hline 
       $ \phantom{\dfrac{x}{2}} J^{\ccR_\ccj}  $ & $-$  &  \checkmark   & $-$  & \checkmark   & \checkmark & $-$& \checkmark & \checkmark 
    \\[1mm] \hline 
   $ \phantom{\dfrac{x}{2}} J^\ccj $ & $-$ &  $-$ & \checkmark   &\checkmark & \checkmark & $-$  &  \checkmark & \checkmark 
    \\[1mm]
\hline
\hline
      $ \phantom{\dfrac{x}{2}}\big[J^{\ccR_\ccj}_\text{eik} \big]_\text{real} $  &  $-$ &   \checkmark  &  $-$  & \checkmark   & \checkmark & $-$&\checkmark &$-$ 
    \\[1mm] \hline
   $\phantom{\dfrac{x}{2}} \big[J^\ccj_\text{eik} \big]_\text{real} $  & $-$  &  $-$ & \checkmark  & \checkmark   & \checkmark & $-$&\checkmark &$-$ 
    \\[1mm]
\hline
   $\phantom{\dfrac{x}{2}} J^{\ccR_\ccj}_\text{eik} $  & $-$  & \checkmark  &  $-$  & \checkmark   & \checkmark & $-$&\checkmark &$-$ 
    \\[1mm]
\hline
   $\phantom{\dfrac{x}{2}} J^\ccj_\text{eik} $   & $-$   &  $-$ & \checkmark & \checkmark   & \checkmark & $-$&\checkmark &$-$ 
    \\[1mm]
\hline
\hline
        $ J^{\ccR_\ccj} - J^{\ccR_\ccj}_\text{eik}$ &  $-$ & $-$   &$-$& $-$  & $-$ & $-$&\checkmark  &\checkmark  
\\[1mm]
\hline
       $ J^\ccj - J^\ccj_\text{eik} $  &  $-$ & $-$  &$-$ & $-$   & $-$ & $-$ &\checkmark  &\checkmark  
    \\[1mm] \hline
    \end{tabular} 
    \caption{IR singularities in virtual and real phase-space integrals at order $\alpha_s$. Note that the jet functions and eikonal jet functions
are all IR divergent but the IR divergences cancels in their difference. Indeed, the subtracted jet function is infrared safe and has no dependence on $\omega$ or $\delta_\ccj$.
 Note that $ \log \omega \log R $ and $\log \omega \log \delta_\ccj$ terms describe soft-collinear singularities. After subtracting the eikonal jet function from the inclusive jet function, these
singularities drop out. 
 } \label{IRsingular}
\end{center} 
\end{table}

\subsection{Soft functions}
Here we focus on the soft function for thrust, defined as
\be
S(\tau) = \sum_{ X_{\scs}
} \int d \Pi_{X_{\scs}} 
\left| \bra{X_{\scs}} Y_\ccO^\dg \cdots Y_\rN \ket{0} \right|^2
\delta \Big(\tau- \frac{1}{2Q}\Omega_\tau(p_{X_{\scs}})\Big)
\ee
with $\Omega_\tau(p)$ the measurement function for thrust defined in Eq.~\eqref{Omegadef} and $X_\scs$ the appropriately phase-space restricted set of states.

The virtual contribution to the soft function comes from the matrix element $\bra{0} Y_\ccO^\dg Y_\ccT\ket{0}$ and is independent of phase-space
restrictions. To order $\alpha_s$,
\be
\bra{0} Y_\ccO^\dg Y_\ccT \ket{0} =1+  g_s^2 \mu^{2 \e} C_F  \int \frac{ \rd^{d} k }{ (2 \pi)^d }   \, 2 \left( \frac{1}{- k^+ + Q\omega } \frac{1}{k^- + Q\omega }   \right)  \frac{1}{k^2} 
\ee
so that
\begin{align}
\label{VirtSoftInc}
\big[ S(\tau) \big]_\text{virt.} 
&= -  \delta(\tau)  \frac{ \alpha_s C_F}{ \pi} \frac{1}{\Gamma(1-\e)} \left( \frac{4 \pi \mu^2}{ Q^2} \right)^\e  \, (-\omega)^{- \e} (\omega)^{-\e} \Big( \Gamma(\e) \Gamma(1-\e) \Big)^2 \\
&=   \delta(\tau) \frac{ \alpha_s C_F}{ \pi} \left( \frac{\wtd \mu^2}{ Q^2} \right)^\e     \left( -\frac{1}{\e^2} + \frac{2}{\e} \ln \omega - \frac{\pi^2}{4} - 2 \ln^2 \omega \right)
\end{align}
We see that the virtual contribution has double poles in the UV and IR. In pure dimensional regularization, these double poles would exactly cancel and
the virtual contribution exactly vanishes. 

Next, we compute the real-emission contributions. First, we consider the inclusive soft function.
At order $\alpha_s$, the real emission contribution is given
\begin{multline}
 \big[S(\tau) \big]_{\text{real}} = 2 g_s^2 \mu^{2 \e} C_F  \int \frac{ \rd^{d} k }{ (2 \pi)^d }   2 \left( \frac{1}{k^+ + Q\omega } \frac{1}{k^- +Q\omega }   \right)  2 \pi \delta(k^+ k^- - k^2_\perp) \theta (k^+) \theta (k^-) 
\\
\times \left[\theta(k^- - k^+) \delta(\tau -\frac{ k^+ }{Q} ) + \theta(k^+ - k^-) \delta(\tau - \frac{k^-}{Q}) \right]
\quad
\end{multline}
As with the jet function, we write the result as
\be
\big[ S \big]_{\text{real}}
=  \frac{ \alpha_s C_F}{2 \pi} \frac{1}{\Gamma(1-\e)} \left( \frac{4 \pi \mu^2}{ Q^2} \right)^\e 
 \frac{1}{\tau+\omega}\frac{1}{\tau^\e} \, \cI_{\scs} (\tau)
\ee
We will use this parameterization for the inclusive and phase-space restricted soft functions.

For the inclusive soft function, we find
\be
\cI_\scs (\tau) = \int_\tau^\infty \rd x  \, \frac{2}{x+ \omega}  \left[ \frac{1}{x} \right]^\e  
= \omega^{-\e} B\left(  \frac{\omega}{ \tau}; \e, 0 \right)  = \frac{2}{\e} - 2 \ln \left( \tau + \omega \right) + \cO(\e)
\ee
so that, using Eq.~\eqref{Itodist},
\be
 \label{RealSoftInc}
\big[ S (\tau) \big]_\text{real} 
 =\frac{ \alpha_s C_F}{ \pi}  \left( \frac{ \wtd \mu^2}{ Q^2} \right)^\e   \left\{ \delta(\tau) \left( - \frac{2}{\e} \ln \omega + \frac{\pi^2}{3} + 2 \ln^2 \omega \right)  +  \left[  \frac{1}{\tau } \left( \frac{2}{\e} - 4 \ln \tau \right)  \right]_+   \right\}
\ee
adding the virtual contribution, we get to order $\alpha_s$
\be
S(\tau) = \delta(\tau) + \frac{ \alpha_s C_F}{ \pi}  \left( \frac{ \wtd \mu^2}{ Q^2} \right)^\e   \left\{ 
\delta(\tau) \left(- \frac{1}{\e^2}  + \frac{\pi^2}{12} \right)  +  \left[  \frac{1}{\tau } \left( \frac{2}{\e} - 4 \ln \tau \right)  \right]_+   \right\}
\ee
which is IR finite (no dependence on $\omega$). The UV poles are removed by renormalization.

The inclusive soft function is IR finite, since all $\omega$ divergences  cancel between real and  virtual diagrams, as we can see in  \Eq{RealSoftInc} and \Eq{VirtSoftInc}. 
As with the jet functions, $\Theta_\text{\red{Res}}$, represents the phase-space restriction. For the soft function in the ball, the phase space restriction is $\Theta_\scL \equiv  \theta ( \Lambda - k^- ) \theta (k^- -k^+ ) + \theta (\Lambda - k^+ ) \theta (k^+ - k^- ) $; for the soft function outside the cones,  $\Theta_\scRb \equiv \theta ( R-  k^-/ k^+ ) \theta ( k^-/k^+ - 1/R ) $; finally, for the soft function in the ball and outside the cones, $\Theta_{\scL \scRb} \equiv \Theta_\scL \Theta_\scRb  $.

For the soft function phase-space restricted to radiation outside of cones around the jet axes (but with no cutoff $\Lambda$ on energy), we find
\be
\cI_\scs^\scRb (\tau) = \int_\tau^{\tau/R} \rd x  \, \frac{2}{x+ \omega}  \left[ \frac{1}{x} \right]^\e  
= 2 \ln \Big( \frac{\tau}{R} + \omega \Big) - 2 \ln \big( \tau + \omega \big) + \cO(\e)
\ee
Note that this soft function is not UV divergent, since at finite $\tau$ there is an implicit energy cutoff $E < \frac{\tau}{2} (1+\frac{1}{R})$, as discussed 
below Eq.~\eqref{softresExpResultR}. 
It leads to
\begin{multline}
 \label{SRbar}
\Big[S^{ \scRb} (\tau) \Big]_\text{real} 
 =\frac{ \alpha_s C_F}{ \pi}    \Bigg\{ \delta(\tau)
 \left[  2 \ln \omega \ln \Big(\frac{R}{ -1+ R} \Big)  - \ln^2 \omega -2 \text{Li}_2 \left( \frac{1+\omega}{ \omega( 1- R)} \right)+ 2 \text{Li}_2 \left( \frac{1}{1-R} \right)  \right]\\
  +  \left[  \frac{2}{\tau } \ln R  \right]_+   \Bigg\}
\\
=\frac{ \alpha_s C_F}{ \pi}    \left\{ \delta(\tau) \left( - \frac{ \pi^2}{3} + 2\ln \omega \ln R + \cO(R,\omega) \right)  +  \left[  \frac{2}{\tau } \ln R  \right]_+   \right\}\hspace{2cm}
\end{multline}
Note that the IR-divergent $\ln \omega \ln R$ term does {\it not} cancel against the virtual correction. Thus the soft function outside the cones is not infrared
safe.

For the soft function phase-space restricted to a ball with energy less than $\Lambda$, but no angular restriction, we find
\be
\cI_\scs^\scL  (\tau) = \int_\tau^{\Lambda/Q} \rd x  \, \frac{2}{x + \omega}  \left[ \frac{1}{x} \right]^\e   = 2 \ln \Big( \frac{\Lambda}{Q} + \omega \Big) - 2 \ln (\tau + \omega)
\ee
This is also UV finite, as expected. To get the soft function outside the cones with energy less then $\Lambda$, we can simply combine these
two expressions with $\theta$-functions:
\be
\cI_\scs^{\scL \scRb}  (\tau) = \cI_\scs^\scL \, \theta \Big( \frac{\tau}{ R} - \Lambda \Big) + \cI_\scs^\scRb \,  \theta \Big( \Lambda- \frac{\tau}{ R} \Big)
\ee
which leads to
\begin{multline}
\Big[S^{ \scL \scRb  } (\tau) \Big]_\text{real} =  \delta(\tau) + C_F \frac{ \alpha_s }{ \pi}  \left( \frac{\mu^2}{ Q^2} \right)^\e \Bigg\{ 
  \delta(\tau) \left[ - \frac{  \pi^2}{3} + 2 \ln \omega \ln R + \cO(R)  \right] 
 -  \bigg[\frac{2 }{\tau}  \ln R  \bigg]_ + \Bigg\}  \theta \Big(\Lambda - \frac{\tau}{R} \Big)
\label{softresExpResult}\\ 
-  C_F \frac{ \alpha_s }{ \pi}  \left( \frac{\mu^2}{ Q^2} \right)^\e \Bigg\{   \delta(\tau) \left[  \frac{1}{\e^2} + \frac{ \pi^2}{4}  + 2 \left(\frac{1}{\e} \ln \omega + \ln \omega \ln \frac{\Lambda}{Q} -  \frac{1}{2} \ln^2 \omega \right) \right]
 +  \bigg[\frac{2 }{\tau }  \ln \frac{\tau Q}{\Lambda} \bigg]_+   \Bigg\} \theta\Big( \frac{\tau}{R} - \Lambda \Big)  
\end{multline} 

Combining this with the virtual contribution leads to Eq.~\eqref{SRL}.
Note that for $\tau$ sufficiently small ($\tau   < \Lambda R$),  $\cI_\scs^{\scL \scRb}$ and hence $S^{\scL \scRb}$ does not depend on $\Lambda$. 
In particular, at leading power in $\tau$ we have
\be
S^{\scL \scRb} \simeq S^{(\scL= \infty) \scRb} 
\ee
as in \Eq{InfiniteLambda}.

 To repeat, we find that the inclusive soft function is IR finite while $S^{\scRb}$ and $S^{\scL\scRb}$ are not, due to an incomplete cancellation  of IR singularities between virtual and real contributions.  Indeed, $S^{\scRb}$ contains a $\ln \omega \ln R$ term,  coming from the real-soft region, and
a $\ln^2 \omega$ term, coming from the virtual-soft-collinear region. These singular terms exactly  match with those in $J_\text{eik}^{\ccR_\ccj}$, according to \Eq{SRbar} and \Eq{JeikRj}.

\bibliography{phasespace}

\providecommand{\href}[2]{#2}\begingroup\raggedright\begin{thebibliography}{10}

\bibitem{Dasgupta:2001sh}
M.~Dasgupta and G.~Salam, ``{Resummation of nonglobal QCD observables},''
  \href{http://dx.doi.org/10.1016/S0370-2693(01)00725-0}{{\em Phys.Lett.}
  {\bfseries B512} (2001) 323--330},
\href{http://arxiv.org/abs/hep-ph/0104277}{{\ttfamily arXiv:hep-ph/0104277
  [hep-ph]}}.

\bibitem{Banfi:2002hw}
A.~Banfi, G.~Marchesini, and G.~Smye, ``{Away from jet energy flow},''
  \href{http://dx.doi.org/10.1088/1126-6708/2002/08/006}{{\em JHEP} {\bfseries
  0208} (2002) 006},
\href{http://arxiv.org/abs/hep-ph/0206076}{{\ttfamily arXiv:hep-ph/0206076
  [hep-ph]}}.

\bibitem{Kelley:2011tj}
R.~Kelley, M.~D. Schwartz, and H.~X. Zhu, ``{Resummation of jet mass with and
  without a jet veto},''
\href{http://arxiv.org/abs/1102.0561}{{\ttfamily arXiv:1102.0561 [hep-ph]}}.

\bibitem{Schwartz:2014wha}
M.~D. Schwartz and H.~X. Zhu, ``{Nonglobal logarithms at three loops, four
  loops, five loops, and beyond},''
  \href{http://dx.doi.org/10.1103/PhysRevD.90.065004}{{\em Phys.Rev.}
  {\bfseries D90} no.~6, (2014) 065004},
\href{http://arxiv.org/abs/1403.4949}{{\ttfamily arXiv:1403.4949 [hep-ph]}}.

\bibitem{Larkoski:2015zka}
A.~J. Larkoski, I.~Moult, and D.~Neill, ``{Non-Global Logarithms,
  Factorization, and the Soft Substructure of Jets},''
\href{http://arxiv.org/abs/1501.04596}{{\ttfamily arXiv:1501.04596 [hep-ph]}}.

\bibitem{Caron-Huot:2015bja}
S.~Caron-Huot, ``{Resummation of non-global logarithms and the BFKL
  equation},''
\href{http://arxiv.org/abs/1501.03754}{{\ttfamily arXiv:1501.03754 [hep-ph]}}.

\bibitem{Khelifa-Kerfa:2015mma}
K.~Khelifa-Kerfa and Y.~Delenda, ``{Non-global logarithms at finite Nc beyond
  leading order},''
\href{http://arxiv.org/abs/1501.00475}{{\ttfamily arXiv:1501.00475 [hep-ph]}}.

\bibitem{Feige:2013zla}
I.~Feige and M.~D. Schwartz, ``{An on-shell approach to factorization},''
  \href{http://dx.doi.org/10.1103/PhysRevD.88.065021}{{\em Phys.Rev.}
  {\bfseries D88} no.~6, (2013) 065021},
\href{http://arxiv.org/abs/1306.6341}{{\ttfamily arXiv:1306.6341 [hep-th]}}.

\bibitem{Feige:2014wja}
I.~Feige and M.~D. Schwartz, ``{Hard-Soft-Collinear Factorization to All
  Orders},'' \href{http://dx.doi.org/10.1103/PhysRevD.90.105020}{{\em
  Phys.Rev.} {\bfseries D90} no.~10, (2014) 105020},
\href{http://arxiv.org/abs/1403.6472}{{\ttfamily arXiv:1403.6472 [hep-ph]}}.

\bibitem{Bloch:1937pw}
F.~Bloch and A.~Nordsieck, ``{Note on the Radiation Field of the electron},''
\href{http://dx.doi.org/10.1103/PhysRev.52.54}{{\em Phys.Rev.} {\bfseries 52}
  (1937) 54--59}.

\bibitem{Weinberg:1964ew}
S.~Weinberg, ``{Photons and Gravitons in s Matrix Theory: Derivation of Charge
  Conservation and Equality of Gravitational and Inertial Mass},''
\href{http://dx.doi.org/10.1103/PhysRev.135.B1049}{{\em Phys.Rev.} {\bfseries
  135} (1964) B1049--B1056}.

\bibitem{Coleman:1965xm}
S.~Coleman and R.~Norton, ``{Singularities in the physical region},''
\href{http://dx.doi.org/10.1007/BF02750472}{{\em Nuovo Cim.} {\bfseries 38}
  (1965) 438--442}.

\bibitem{Sterman:1978bi}
G.~F. Sterman, ``{Mass Divergences in Annihilation Processes. 1. Origin and
  Nature of Divergences in Cut Vacuum Polarization Diagrams},''
\href{http://dx.doi.org/10.1103/PhysRevD.17.2773}{{\em Phys.Rev.} {\bfseries
  D17} (1978) 2773}.

\bibitem{Collins:1981uk}
J.~C. Collins and D.~E. Soper, ``{Back-To-Back Jets in QCD},''
\href{http://dx.doi.org/10.1016/0550-3213(81)90339-4}{{\em Nucl.Phys.}
  {\bfseries B193} (1981) 381}.

\bibitem{Bauer:2001yt}
C.~W. Bauer, D.~Pirjol, and I.~W. Stewart, ``{Soft collinear factorization in
  effective field theory},''
  \href{http://dx.doi.org/10.1103/PhysRevD.65.054022}{{\em Phys.Rev.}
  {\bfseries D65} (2002) 054022},
\href{http://arxiv.org/abs/hep-ph/0109045}{{\ttfamily arXiv:hep-ph/0109045
  [hep-ph]}}.

\bibitem{Freedman:2011kj}
S.~M. Freedman and M.~Luke, ``{SCET, QCD and Wilson Lines},''
  \href{http://dx.doi.org/10.1103/PhysRevD.85.014003}{{\em Phys.Rev.}
  {\bfseries D85} (2012) 014003},
\href{http://arxiv.org/abs/1107.5823}{{\ttfamily arXiv:1107.5823 [hep-ph]}}.

\bibitem{Akhoury:1998gs}
R.~Akhoury, M.~G. Sotiropoulos, and G.~F. Sterman, ``{An Operator expansion for
  the elastic limit},''
  \href{http://dx.doi.org/10.1103/PhysRevLett.81.3819}{{\em Phys.Rev.Lett.}
  {\bfseries 81} (1998) 3819--3822},
\href{http://arxiv.org/abs/hep-ph/9807330}{{\ttfamily arXiv:hep-ph/9807330
  [hep-ph]}}.

\bibitem{Collins:1989bt}
J.~C. Collins, ``{Sudakov form-factors},'' {\em Adv.Ser.Direct.High Energy
  Phys.} {\bfseries 5} (1989) 573--614,
\href{http://arxiv.org/abs/hep-ph/0312336}{{\ttfamily arXiv:hep-ph/0312336
  [hep-ph]}}.

\bibitem{Collins:1989gx}
J.~C. Collins, D.~E. Soper, and G.~F. Sterman, ``{Factorization of Hard
  Processes in QCD},'' {\em Adv.Ser.Direct.High Energy Phys.} {\bfseries 5}
  (1988) 1--91,
\href{http://arxiv.org/abs/hep-ph/0409313}{{\ttfamily arXiv:hep-ph/0409313
  [hep-ph]}}.

\bibitem{Berger:2003iw}
C.~F. Berger, T.~Kucs, and G.~F. Sterman, ``{Event shape / energy flow
  correlations},'' \href{http://dx.doi.org/10.1103/PhysRevD.68.014012}{{\em
  Phys.Rev.} {\bfseries D68} (2003) 014012},
\href{http://arxiv.org/abs/hep-ph/0303051}{{\ttfamily arXiv:hep-ph/0303051
  [hep-ph]}}.

\bibitem{Manohar:2006nz}
A.~V. Manohar and I.~W. Stewart, ``{The Zero-Bin and Mode Factorization in
  Quantum Field Theory},''
  \href{http://dx.doi.org/10.1103/PhysRevD.76.074002}{{\em Phys.Rev.}
  {\bfseries D76} (2007) 074002},
\href{http://arxiv.org/abs/hep-ph/0605001}{{\ttfamily arXiv:hep-ph/0605001
  [hep-ph]}}.

\bibitem{Lee:2006nr}
C.~Lee and G.~F. Sterman, ``{Momentum Flow Correlations from Event Shapes:
  Factorized Soft Gluons and Soft-Collinear Effective Theory},''
  \href{http://dx.doi.org/10.1103/PhysRevD.75.014022}{{\em Phys.Rev.}
  {\bfseries D75} (2007) 014022},
\href{http://arxiv.org/abs/hep-ph/0611061}{{\ttfamily arXiv:hep-ph/0611061
  [hep-ph]}}.

\bibitem{Idilbi:2007ff}
A.~Idilbi and T.~Mehen, ``{On the equivalence of soft and zero-bin
  subtractions},'' \href{http://dx.doi.org/10.1103/PhysRevD.75.114017}{{\em
  Phys.Rev.} {\bfseries D75} (2007) 114017},
\href{http://arxiv.org/abs/hep-ph/0702022}{{\ttfamily arXiv:hep-ph/0702022
  [HEP-PH]}}.

\bibitem{Idilbi:2007yi}
A.~Idilbi and T.~Mehen, ``{Demonstration of the equivalence of soft and
  zero-bin subtractions},''
  \href{http://dx.doi.org/10.1103/PhysRevD.76.094015}{{\em Phys.Rev.}
  {\bfseries D76} (2007) 094015},
\href{http://arxiv.org/abs/0707.1101}{{\ttfamily arXiv:0707.1101 [hep-ph]}}.

\bibitem{Beneke:1997zp}
M.~Beneke and V.~A. Smirnov, ``{Asymptotic expansion of Feynman integrals near
  threshold},'' \href{http://dx.doi.org/10.1016/S0550-3213(98)00138-2}{{\em
  Nucl.Phys.} {\bfseries B522} (1998) 321--344},
\href{http://arxiv.org/abs/hep-ph/9711391}{{\ttfamily arXiv:hep-ph/9711391
  [hep-ph]}}.

\bibitem{Beneke:2002ph}
M.~Beneke, A.~Chapovsky, M.~Diehl, and T.~Feldmann, ``{Soft collinear effective
  theory and heavy to light currents beyond leading power},''
  \href{http://dx.doi.org/10.1016/S0550-3213(02)00687-9}{{\em Nucl.Phys.}
  {\bfseries B643} (2002) 431--476},
\href{http://arxiv.org/abs/hep-ph/0206152}{{\ttfamily arXiv:hep-ph/0206152
  [hep-ph]}}.

\bibitem{Beneke:2002ni}
M.~Beneke and T.~Feldmann, ``{Multipole expanded soft collinear effective
  theory with nonAbelian gauge symmetry},''
  \href{http://dx.doi.org/10.1016/S0370-2693(02)03204-5}{{\em Phys.Lett.}
  {\bfseries B553} (2003) 267--276},
\href{http://arxiv.org/abs/hep-ph/0211358}{{\ttfamily arXiv:hep-ph/0211358
  [hep-ph]}}.

\bibitem{Becher:2014oda}
T.~Becher, A.~Broggio, and A.~Ferroglia, ``{Introduction to Soft-Collinear
  Effective Theory},''
\href{http://arxiv.org/abs/1410.1892}{{\ttfamily arXiv:1410.1892 [hep-ph]}}.

\bibitem{Bauer:2000yr}
C.~W. Bauer, S.~Fleming, D.~Pirjol, and I.~W. Stewart, ``{An Effective field
  theory for collinear and soft gluons: Heavy to light decays},''
  \href{http://dx.doi.org/10.1103/PhysRevD.63.114020}{{\em Phys.Rev.}
  {\bfseries D63} (2001) 114020},
\href{http://arxiv.org/abs/hep-ph/0011336}{{\ttfamily arXiv:hep-ph/0011336
  [hep-ph]}}.

\bibitem{Bauer:2002nz}
C.~W. Bauer, S.~Fleming, D.~Pirjol, I.~Z. Rothstein, and I.~W. Stewart, ``{Hard
  scattering factorization from effective field theory},''
  \href{http://dx.doi.org/10.1103/PhysRevD.66.014017}{{\em Phys.Rev.}
  {\bfseries D66} (2002) 014017},
\href{http://arxiv.org/abs/hep-ph/0202088}{{\ttfamily arXiv:hep-ph/0202088
  [hep-ph]}}.

\bibitem{Alioli:2013hba}
S.~Alioli and J.~R. Walsh, ``{Jet Veto Clustering Logarithms Beyond Leading
  Order},'' \href{http://dx.doi.org/10.1007/JHEP03(2014)119}{{\em JHEP}
  {\bfseries 1403} (2014) 119},
\href{http://arxiv.org/abs/1311.5234}{{\ttfamily arXiv:1311.5234 [hep-ph]}}.

\bibitem{Frixione:1995ms}
S.~Frixione, Z.~Kunszt, and A.~Signer, ``{Three jet cross-sections to
  next-to-leading order},''
  \href{http://dx.doi.org/10.1016/0550-3213(96)00110-1}{{\em Nucl.Phys.}
  {\bfseries B467} (1996) 399--442},
\href{http://arxiv.org/abs/hep-ph/9512328}{{\ttfamily arXiv:hep-ph/9512328
  [hep-ph]}}.

\bibitem{Kunszt:1992tn}
Z.~Kunszt and D.~E. Soper, ``{Calculation of jet cross-sections in hadron
  collisions at order alpha-s**3},''
\href{http://dx.doi.org/10.1103/PhysRevD.46.192}{{\em Phys.Rev.} {\bfseries
  D46} (1992) 192--221}.

\bibitem{Catani:1996vz}
S.~Catani and M.~Seymour, ``{A General algorithm for calculating jet
  cross-sections in NLO QCD},''
  \href{http://dx.doi.org/10.1016/S0550-3213(96)00589-5}{{\em Nucl.Phys.}
  {\bfseries B485} (1997) 291--419},
\href{http://arxiv.org/abs/hep-ph/9605323}{{\ttfamily arXiv:hep-ph/9605323
  [hep-ph]}}.

\bibitem{GehrmannDeRidder:2005cm}
A.~Gehrmann-De~Ridder, T.~Gehrmann, and E.~N. Glover, ``{Antenna subtraction at
  NNLO},'' \href{http://dx.doi.org/10.1088/1126-6708/2005/09/056}{{\em JHEP}
  {\bfseries 0509} (2005) 056},
\href{http://arxiv.org/abs/hep-ph/0505111}{{\ttfamily arXiv:hep-ph/0505111
  [hep-ph]}}.

\bibitem{Catani:1992ua}
S.~Catani, L.~Trentadue, G.~Turnock, and B.~Webber, ``{Resummation of large
  logarithms in e+ e- event shape distributions},''
\href{http://dx.doi.org/10.1016/0550-3213(93)90271-P}{{\em Nucl.Phys.}
  {\bfseries B407} (1993) 3--42}.

\bibitem{Schwartz:2007ib}
M.~D. Schwartz, ``{Resummation and NLO matching of event shapes with effective
  field theory},'' \href{http://dx.doi.org/10.1103/PhysRevD.77.014026}{{\em
  Phys.Rev.} {\bfseries D77} (2008) 014026},
\href{http://arxiv.org/abs/0709.2709}{{\ttfamily arXiv:0709.2709 [hep-ph]}}.

\bibitem{Becher:2008cf}
T.~Becher and M.~D. Schwartz, ``{A precise determination of $\alpha_s$ from LEP
  thrust data using effective field theory},''
  \href{http://dx.doi.org/10.1088/1126-6708/2008/07/034}{{\em JHEP} {\bfseries
  0807} (2008) 034},
\href{http://arxiv.org/abs/0803.0342}{{\ttfamily arXiv:0803.0342 [hep-ph]}}.

\bibitem{Abbate:2010xh}
R.~Abbate, M.~Fickinger, A.~H. Hoang, V.~Mateu, and I.~W. Stewart, ``{Thrust at
  $N^3LL$ with Power Corrections and a Precision Global Fit for alphas(mZ)},''
  \href{http://dx.doi.org/10.1103/PhysRevD.83.074021}{{\em Phys.Rev.}
  {\bfseries D83} (2011) 074021},
\href{http://arxiv.org/abs/1006.3080}{{\ttfamily arXiv:1006.3080 [hep-ph]}}.

\bibitem{Berger:2003pk}
C.~F. Berger and G.~F. Sterman, ``{Scaling rule for nonperturbative radiation
  in a class of event shapes},''
  \href{http://dx.doi.org/10.1088/1126-6708/2003/09/058}{{\em JHEP} {\bfseries
  0309} (2003) 058},
\href{http://arxiv.org/abs/hep-ph/0307394}{{\ttfamily arXiv:hep-ph/0307394
  [hep-ph]}}.

\bibitem{Hornig:2009vb}
A.~Hornig, C.~Lee, and G.~Ovanesyan, ``{Effective Predictions of Event Shapes:
  Factorized, Resummed, and Gapped Angularity Distributions},''
  \href{http://dx.doi.org/10.1088/1126-6708/2009/05/122}{{\em JHEP} {\bfseries
  0905} (2009) 122},
\href{http://arxiv.org/abs/0901.3780}{{\ttfamily arXiv:0901.3780 [hep-ph]}}.

\bibitem{Chien:2010kc}
Y.-T. Chien and M.~D. Schwartz, ``{Resummation of heavy jet mass and comparison
  to LEP data},'' \href{http://dx.doi.org/10.1007/JHEP08(2010)058}{{\em JHEP}
  {\bfseries 1008} (2010) 058},
\href{http://arxiv.org/abs/1005.1644}{{\ttfamily arXiv:1005.1644 [hep-ph]}}.

\bibitem{Parisi:1978eg}
G.~Parisi, ``{Super Inclusive Cross-Sections},''
\href{http://dx.doi.org/10.1016/0370-2693(78)90061-8}{{\em Phys.Lett.}
  {\bfseries B74} (1978) 65}.

\bibitem{Donoghue:1979vi}
J.~F. Donoghue, F.~Low, and S.-Y. Pi, ``{Tensor Analysis of Hadronic Jets in
  Quantum Chromodynamics},''
\href{http://dx.doi.org/10.1103/PhysRevD.20.2759}{{\em Phys.Rev.} {\bfseries
  D20} (1979) 2759}.

\bibitem{Hoang:2015hka}
A.~H. Hoang, D.~W. Kolodrubetz, V.~Mateu, and I.~W. Stewart, ``{A Precise
  Determination of $\alpha_s$ from the C-parameter Distribution},''
\href{http://arxiv.org/abs/1501.04111}{{\ttfamily arXiv:1501.04111 [hep-ph]}}.

\bibitem{Rakow:1981qn}
P.~E. Rakow and B.~Webber, ``{Transverse Momentum Moments of Hadron
  Distributions in {QCD} Jets},''
\href{http://dx.doi.org/10.1016/0550-3213(81)90286-8}{{\em Nucl.Phys.}
  {\bfseries B191} (1981) 63}.

\bibitem{Ellis:1986ig}
R.~K. Ellis and B.~Webber, ``{QCD Jet Broadening in Hadron Hadron
  Collisions},''
{\em Conf.Proc.} {\bfseries C860623} (1986) 74.

\bibitem{Catani:1992jc}
S.~Catani, G.~Turnock, and B.~Webber, ``{Jet broadening measures in $e^{+}
  e^{-}$ annihilation},''
\href{http://dx.doi.org/10.1016/0370-2693(92)91565-Q}{{\em Phys.Lett.}
  {\bfseries B295} (1992) 269--276}.

\bibitem{Chiu:2011qc}
J.-y. Chiu, A.~Jain, D.~Neill, and I.~Z. Rothstein, ``{The Rapidity
  Renormalization Group},''
  \href{http://dx.doi.org/10.1103/PhysRevLett.108.151601}{{\em Phys.Rev.Lett.}
  {\bfseries 108} (2012) 151601},
\href{http://arxiv.org/abs/1104.0881}{{\ttfamily arXiv:1104.0881 [hep-ph]}}.

\bibitem{Becher:2011pf}
T.~Becher, G.~Bell, and M.~Neubert, ``{Factorization and Resummation for Jet
  Broadening},'' \href{http://dx.doi.org/10.1016/j.physletb.2011.09.005}{{\em
  Phys.Lett.} {\bfseries B704} (2011) 276--283},
\href{http://arxiv.org/abs/1104.4108}{{\ttfamily arXiv:1104.4108 [hep-ph]}}.

\bibitem{Laenen:1998qw}
E.~Laenen, G.~Oderda, and G.~F. Sterman, ``{Resummation of threshold
  corrections for single particle inclusive cross-sections},''
  \href{http://dx.doi.org/10.1016/S0370-2693(98)00960-5}{{\em Phys.Lett.}
  {\bfseries B438} (1998) 173--183},
\href{http://arxiv.org/abs/hep-ph/9806467}{{\ttfamily arXiv:hep-ph/9806467
  [hep-ph]}}.

\bibitem{Becher:2007ty}
T.~Becher, M.~Neubert, and G.~Xu, ``{Dynamical Threshold Enhancement and
  Resummation in Drell-Yan Production},''
  \href{http://dx.doi.org/10.1088/1126-6708/2008/07/030}{{\em JHEP} {\bfseries
  0807} (2008) 030},
\href{http://arxiv.org/abs/0710.0680}{{\ttfamily arXiv:0710.0680 [hep-ph]}}.

\bibitem{Manohar:2003vb}
A.~V. Manohar, ``{Deep inelastic scattering as x -> 1 using soft collinear
  effective theory},'' \href{http://dx.doi.org/10.1103/PhysRevD.68.114019}{{\em
  Phys.Rev.} {\bfseries D68} (2003) 114019},
\href{http://arxiv.org/abs/hep-ph/0309176}{{\ttfamily arXiv:hep-ph/0309176
  [hep-ph]}}.

\bibitem{Becher:2009th}
T.~Becher and M.~D. Schwartz, ``{Direct photon production with effective field
  theory},'' \href{http://dx.doi.org/10.1007/JHEP02(2010)040}{{\em JHEP}
  {\bfseries 1002} (2010) 040},
\href{http://arxiv.org/abs/0911.0681}{{\ttfamily arXiv:0911.0681 [hep-ph]}}.

\bibitem{Kidonakis:2003bh}
N.~Kidonakis and J.~Owens, ``{Next-to-next-to-leading order soft gluon
  corrections in direct photon production},''
  \href{http://dx.doi.org/10.1142/S0217751X04017458}{{\em Int.J.Mod.Phys.}
  {\bfseries A19} (2004) 149--158},
\href{http://arxiv.org/abs/hep-ph/0307352}{{\ttfamily arXiv:hep-ph/0307352
  [hep-ph]}}.

\bibitem{Gonsalves:2005ng}
R.~J. Gonsalves, N.~Kidonakis, and A.~Sabio~Vera, ``{$W$ production at large
  transverse momentum at the large hadron collider},''
  \href{http://dx.doi.org/10.1103/PhysRevLett.95.222001}{{\em Phys.Rev.Lett.}
  {\bfseries 95} (2005) 222001},
\href{http://arxiv.org/abs/hep-ph/0507317}{{\ttfamily arXiv:hep-ph/0507317
  [hep-ph]}}.

\bibitem{Becher:2011fc}
T.~Becher, C.~Lorentzen, and M.~D. Schwartz, ``{Resummation for W and Z
  production at large pT},''
  \href{http://dx.doi.org/10.1103/PhysRevLett.108.012001}{{\em Phys.Rev.Lett.}
  {\bfseries 108} (2012) 012001},
\href{http://arxiv.org/abs/1106.4310}{{\ttfamily arXiv:1106.4310 [hep-ph]}}.

\bibitem{Becher:2012xr}
T.~Becher, C.~Lorentzen, and M.~D. Schwartz, ``{Precision Direct Photon and
  W-Boson Spectra at High $p_T$ and Comparison to LHC Data},''
  \href{http://dx.doi.org/10.1103/PhysRevD.86.054026}{{\em Phys.Rev.}
  {\bfseries D86} (2012) 054026},
\href{http://arxiv.org/abs/1206.6115}{{\ttfamily arXiv:1206.6115 [hep-ph]}}.

\bibitem{Chien:2012ur}
Y.-T. Chien, R.~Kelley, M.~D. Schwartz, and H.~X. Zhu, ``{Resummation of Jet
  Mass at Hadron Colliders},''
  \href{http://dx.doi.org/10.1103/PhysRevD.87.014010}{{\em Phys.Rev.}
  {\bfseries D87} (2013) 014010},
\href{http://arxiv.org/abs/1208.0010}{{\ttfamily arXiv:1208.0010}}.

\bibitem{Dasgupta:2012hg}
M.~Dasgupta, K.~Khelifa-Kerfa, S.~Marzani, and M.~Spannowsky, ``{On jet mass
  distributions in Z+jet and dijet processes at the LHC},''
  \href{http://dx.doi.org/10.1007/JHEP10(2012)126}{{\em JHEP} {\bfseries 1210}
  (2012) 126},
\href{http://arxiv.org/abs/1207.1640}{{\ttfamily arXiv:1207.1640 [hep-ph]}}.

\bibitem{Ahrens:2010zv}
V.~Ahrens, A.~Ferroglia, M.~Neubert, B.~D. Pecjak, and L.~L. Yang,
  ``{Renormalization-Group Improved Predictions for Top-Quark Pair Production
  at Hadron Colliders},'' \href{http://dx.doi.org/10.1007/JHEP09(2010)097}{{\em
  JHEP} {\bfseries 1009} (2010) 097},
\href{http://arxiv.org/abs/1003.5827}{{\ttfamily arXiv:1003.5827 [hep-ph]}}.

\bibitem{Kidonakis:2012rm}
N.~Kidonakis, ``{NNLL threshold resummation for top-pair and single-top
  production},'' \href{http://dx.doi.org/10.1134/S1063779614040091}{{\em
  Phys.Part.Nucl.} {\bfseries 45} no.~4, (2014) 714--722},
\href{http://arxiv.org/abs/1210.7813}{{\ttfamily arXiv:1210.7813 [hep-ph]}}.

\bibitem{Stewart:2010tn}
I.~W. Stewart, F.~J. Tackmann, and W.~J. Waalewijn, ``{N-Jettiness: An
  Inclusive Event Shape to Veto Jets},''
  \href{http://dx.doi.org/10.1103/PhysRevLett.105.092002}{{\em Phys.Rev.Lett.}
  {\bfseries 105} (2010) 092002},
\href{http://arxiv.org/abs/1004.2489}{{\ttfamily arXiv:1004.2489 [hep-ph]}}.

\bibitem{Thaler:2010tr}
J.~Thaler and K.~Van~Tilburg, ``{Identifying Boosted Objects with
  N-subjettiness},'' \href{http://dx.doi.org/10.1007/JHEP03(2011)015}{{\em
  JHEP} {\bfseries 1103} (2011) 015},
\href{http://arxiv.org/abs/1011.2268}{{\ttfamily arXiv:1011.2268 [hep-ph]}}.

\bibitem{Feige:2012vc}
I.~Feige, M.~D. Schwartz, I.~W. Stewart, and J.~Thaler, ``{Precision Jet
  Substructure from Boosted Event Shapes},''
  \href{http://dx.doi.org/10.1103/PhysRevLett.109.092001}{{\em Phys.Rev.Lett.}
  {\bfseries 109} (2012) 092001},
\href{http://arxiv.org/abs/1204.3898}{{\ttfamily arXiv:1204.3898 [hep-ph]}}.

\bibitem{Chiu:2009yx}
J.-y. Chiu, A.~Fuhrer, A.~H. Hoang, R.~Kelley, and A.~V. Manohar,
  ``{Soft-Collinear Factorization and Zero-Bin Subtractions},''
  \href{http://dx.doi.org/10.1103/PhysRevD.79.053007}{{\em Phys.Rev.}
  {\bfseries D79} (2009) 053007},
\href{http://arxiv.org/abs/0901.1332}{{\ttfamily arXiv:0901.1332 [hep-ph]}}.

\end{thebibliography}\endgroup
\bibliographystyle{utphys}

\end{document}